\documentclass[journal, twoside]{IEEEtran}
\usepackage{subcaption}
\usepackage{indentfirst}
\usepackage{graphicx}
\usepackage{amssymb}
\usepackage{amsfonts}
\usepackage{mathrsfs}
\usepackage{leftidx}
\usepackage{color}
\usepackage{amsmath}
\usepackage{arydshln}
\usepackage{amsthm}
\usepackage{ragged2e}
\usepackage{cite}
\usepackage{enumerate}
\usepackage{longtable}
\usepackage{float}
\usepackage{stfloats}
\usepackage{hyperref}
\usepackage{algpseudocode}
\usepackage{algorithm}
\usepackage{makecell}
\usepackage{url}
\usepackage{enumitem}
\usepackage{multirow} 
\usepackage{booktabs}
\theoremstyle{plain}

\theoremstyle{remark}

\usepackage{caption}
\usepackage{makecell}

\newcolumntype{P}[1]{>{\raggedright\arraybackslash\footnotesize}m{#1}}
\newcolumntype{A}[1]{>{\centering\arraybackslash\footnotesize}m{#1}}

\usepackage[table,usenames,dvipsnames]{xcolor}
\definecolor{aa}{RGB}{175,238,238}
\definecolor{bb}{RGB}{255,255,255}

\usepackage{bm}
\usepackage{makecell}

\begin{document}

\title{Secure Intellicise Wireless Network: Agentic AI for Coverless Semantic Steganography Communication}

\author{Rui Meng,~\IEEEmembership{Member,~IEEE,} Song Gao, Bingxuan Xu,
Xiaodong Xu,~\IEEEmembership{Senior Member,~IEEE,}

Jianqiao Chen, Nan Ma,~\IEEEmembership{Member,~IEEE,}
Pei Xiao,~\IEEEmembership{Senior Member,~IEEE,} 

Ping Zhang,~\IEEEmembership{Fellow,~IEEE,} 
and Rahim Tafazolli,~\IEEEmembership{Fellow,~IEEE}

\thanks{
This work was supported in part by the National Key Research and Development Program of China under Grant 2020YFB1806905; in part by the National Natural Science Foundation of China under Grant 62501066 and under Grant U24B20131; in part by the Beijing Municipal Natural Science Foundation under Grant L242012; and in part by the Long Term Science and Technology Plan for Broadcasting, Television, and Online Audiovisual Program under Grant 2025AD0300.
\textit{(Corresponding author: Rui Meng and Xiaodong Xu.)}

Rui Meng, Song Gao, Bingxuan Xu, Xiaodong Xu, Nan Ma, and Ping Zhang are with State Key Laboratory of Networking and Switching Technology, Beijing University of Posts and Telecommunications, Beijing, China (e-mail: buptmengrui@bupt.edu.cn; wkd251292@bupt.edu.cn; xubingxuan@bupt.edu.cn; xuxiaodong@bupt.edu.cn; manan@bupt.edu.cn; pzhang@bupt.edu.cn).

Jianqiao Chen is with the ZGC Institute of Ubiquitous-X Innovation and Applications, and Beijing Key Laboratory of 6G DOICT converged and Cloud-Native Mobile Information Networks, Beijing 100876, China.  (email: jqchen1988@163.com).

Pei Xiao and Rahim Tafazolli are with 5GIC \& 6GIC, Institute for Communication Systems (ICS), University of Surrey, Guildford, GU2 7XH, United Kingdom (email: p.xiao@surrey.ac.uk; r.tafazolli@surrey.ac.uk).

}
}

\maketitle

\begin{abstract}
Semantic Communication (SemCom), leveraging its significant advantages in transmission efficiency and reliability, has emerged as a core technology for constructing future \textit{intellicise (intelligent and concise)} wireless networks. However, intelligent attacks represented by semantic eavesdropping pose severe challenges to the security of SemCom. To address this challenge, Semantic Steganographic Communication (SemSteCom) achieves ``invisible'' encryption by implicitly embedding private semantic information into cover modality carriers. The state-of-the-art study has further introduced generative diffusion models to directly generate stega images without relying on original cover images, effectively enhancing steganographic capacity. Nevertheless, the recovery process of private images is highly dependent on the guidance of private semantic keys, which may be inferred by intelligent eavesdroppers, thereby introducing new security threats. To address this issue, we propose an Agentic AI-driven SemSteCom (AgentSemSteCom) scheme, which includes semantic extraction, digital token controlled reference image generation, coverless steganography, semantic codec, and optional task-oriented enhancement modules. The proposed AgentSemSteCom scheme obviates the need for both cover images and private semantic keys, thereby boosting steganographic capacity while reinforcing transmission security. The simulation results on open-source datasets verify that, AgentSemSteCom achieves better transmission quality and higher security levels than the baseline scheme.
\end{abstract}


\section{Introduction}





The deep integration of Artificial Intelligence (AI) technology and communication is one of the core trends driving the evolution of the sixth generation mobile communication (6G) \cite{meng2026intellicise,yining2024intellicise,meng2025generative}. In this context, the \textit{intellicise (intelligent and concise)} wireless network, rooted and guided by information theory, complex science theory, and system theory, realizes holistic optimization through entropy reduction. Intellicise embodies the dual principles of \textit{endogenous intelligence} and \textit{native simplicity}. Leveraging native intelligence, cognitive reshaping, and adaptive characteristics, it constructs intelligent service ecosystems tailored to diverse communication objects, ultimately enabling a self-evolving, self-optimizing, and self-balancing network \cite{zhang2024intellicise,meng2025image,zhang2026towards}.

Semantic communication (SemCom) plays a pivotal role in the intellicise wireless network. Traditional communication based on Shannon information theory focuses on precise transmission of syntax layer symbols and pursues bit-level lossless transmission \cite{xu2023latent}. SemCom shifts towards the semantic layer, emphasizing the priority of meaning expression and allowing for moderate distortion at the symbolic level while preserving semantics \cite{meng2025semantic}. Numerous scholars have dedicated their research to SemCom and proposed key enabling technologies, such as semantic-based coding \cite{bourtsoulatze2019deep,dai2022nonlinear,teng2025conquering}, semantic-based multiple access \cite{zhang2023model}, semantic importance-based transmission \cite{wang2024feature,lu2025important,cao2025importance,liang2025semantic, tang2024contrastive}, semantic knowledge base \cite{fan2025kgrag}, and generative AI-based SemCom \cite{xu2025semantic, tang2024retrieval}. For instance, Zhang \textit{et al.} \cite{zhang2024experts} explored generative AI towards semantic-aware satellite communication systems for model formulation, demonstrating that generative AI can effectively abstract high-level semantic information to guide communication decisions.

In contrast to traditional communication, SemCom faces heightened security threats \cite{meng2025survey}. It must address not only conventional security risks, such as eavesdropping, jamming, and man-in-the-middle attacks, but also emerging threats introduced by semantic models, such as backdoor attacks and model inversion attacks\cite{getu2025semantic}. Due to the open nature of wireless channels, eavesdropping remains one of the most prevalent attacks in SemCom \cite{zhang2025srec}. Adversaries can leverage specialized receiving equipment and advanced signal processing techniques to intercept transmitted signals \cite{do2025security}. Moreover, if they acquire the decoder parameters deployed at legitimate receivers, they could potentially decode some or all private semantic information, even under degraded channel conditions.

To defend against semantic eavesdropping attacks, researchers have recently proposed several approaches. Among them, encryption remains one of the most established techniques and has been widely adopted in SemCom, including cryptography encryption\cite{tung2023deep}, homomorphic encryption\cite{meng2025secure}, and quantum cryptography\cite{kaewpuang2024cooperative}. Based on  physical-layer security\cite{li2023physical}, researchers employ superposition coding\cite{chen2024nearly}, beamforming \cite{dai2024secure} and reconfigurable intelligent surface \cite{zhao2022semkey} to achieve secure SemCom. Meanwhile, some researchers further consider covert communications\cite{xu2024covert, wang2023multi,liu2025learning}, which aims to conceal the transmission behavior of SemCom from potential semantic eavesdroppers.


While covert communications focus on hiding the communication signals \cite{li2025covert}, recent studies shift attention toward concealing the confidential semantic content itself, leading to the exploration of Semantic Steganography Communication (SemSteCom). Rather than directly encrypting semantic representations, SemSteCom conceals the confidential semantic information within inconspicuous information. Towards text transmission, Long \textit{et al.} \cite{long2025scf} proposed a knowledge graph-guided steganography framework to improve semantic coherence and imperceptibility in linguistic domain. Li \textit{et al.} \cite{li2024multi} further embedded textual semantics into visual semantics to enable secure multi-modal semantic communication. Toward image transmission, Tang \textit{et al.} \cite{tang2025towards} and  Ni \textit{et al.} \cite{ni2025controllable} developed Invertible Neural Network (INN)-based SemSteCom schemes, while Huo \textit{et al.} \cite{huo2025image} introduced a Generative Adversarial Network (GAN)-based framework. Also, Chen \textit{et al.} \cite{chencoding} further proposed a coding-guided semantic jamming strategy based on superposition coding to degrade the eavesdropper’s decoding capability. 

However, the above solutions rely on cover modalities to hide secret information, which introduces limitations in steganographic capacity and security. To address these challenges, Gao \textit{et al.} \cite{gao2025semstediff} introduced a coverless SemSteDiff scheme that embeds secret semantics into the diffusion sampling trajectory, allowing legitimate receivers to recover confidential image using private semantic keys to achieve “invisible encryption”.
However, in some scenarios enabled by intellicise wireless networks, intelligent eavesdroppers could infer the private semantic key to generate the secret image based on the semantic background \cite{tang2026rethinking}.
For example, in the intellicise healthcare scenario, an intelligent semantic eavesdropper aware of a patient's medical history could attempt to reconstruct sensitive diagnostic images by inferring semantic information like ``fracture'' or ``tumor''. 

Against this background, we introduce agentic AI\cite{zhang2024interactive} to autonomously determine the generative path of stego image based on a fixed digital token. Unlike conventional generative AI models that follow a static and predefined generation pipeline, agentic AI dynamically perceives semantic characteristics of the secret content and adaptively selects task-specific generative strategies. Specifically, agentic AI utilizes the digital token as a deterministic seed, generating context-adaptive stego images based on the semantic feature perceived from different secret images. It can effectively avoid the risk of information leakage caused by semantic correlation and repeated transmission of private semantic keys, enhancing the security and adaptability of SemSteCom. The main contributions of this paper are summarized as follows:
\begin{itemize}
\item We propose an Agentic AI-driven SemSteCom (AgentSemSteCom) scheme to realize autonomous defense against intelligent semantic eavesdroppers in intellicise wireless networks. Compared with the state-of-the-art SemSteCom schemes \cite{tang2025towards,ni2025controllable,huo2025image,chencoding}, AgentSemSteCom eliminates the requirement for cover images and private semantic keys, enhancing the steganographic capacity and transmission security. To the best of our knowledge, this is the first work to employ agentic AI for secure intellicise wireless networks.


\item 
We develop an agentic AI-driven coverless semantic steganography framework, where the public semantic key and implicit feature are autonomously coordinated by agentic AI to control the stego image generation. The public semantic key determines the content description while the implicit feature constrains the structural distribution, which enables controllable generation. Also, agentic AI employs an invertible sampling strategy, where coupled latent trajectories are exploited to eliminate approximation error in diffusion inversion, thereby resolving semantic drift and enabling accurate recovery of the secret image at the receiver.
\item We propose a digital token–based security mechanism to protect secret images without relying on private semantic keys. Agentic AI utilizes a user-defined digital token as a deterministic random seed to initialize the diffusion noise for reference image generation, ensuring that only legitimate receivers can achieve invertible recovery using the digital token. Moreover, the digital token is also used to disturb the latent vector with controllable noise during diffusion process, where agentic AI applies a binary perturbation mask to flip the specific position determined by digital token. This dual-stage protection enables agentic AI to coordinate reference image generation and noise perturbation under a unified digital token control, effectively defending against the potential eavesdropping risk caused by the leakage of private semantic keys, which further enhances the security of SemSteCom.
\item 
The simulation results on open-source UniStega dataset \cite{yang2024diffstega} demonstrate the effectiveness of the proposed AgentSemSteCom scheme for both the digital and analog SemCom. Compared with SemSteDiff \cite{gao2025semstediff}, our scheme achieves higher Peak Signal-to-Noise Ratio (PSNR) by 2.69 dB, higher Structural Similarity (SSIM) by 8.88$\%$, lower Mean Squared Error (MSE) by 43.75$\%$, and lower Learned Perceptual Image Patch Similarity (LPIPS) by 5.1$\%$. Additionally, the agentic-AI based legitimate receiver with the correct digital token achieves PSNR of 25.37 dB, while the agentic-AI based eavesdropper with the wrong digital token only reaches PSNR of 18.91 dB.

\end{itemize}

\section{System Model}

\subsection{Network Model}
\begin{figure*}
\centering
\includegraphics[width=0.8\linewidth]{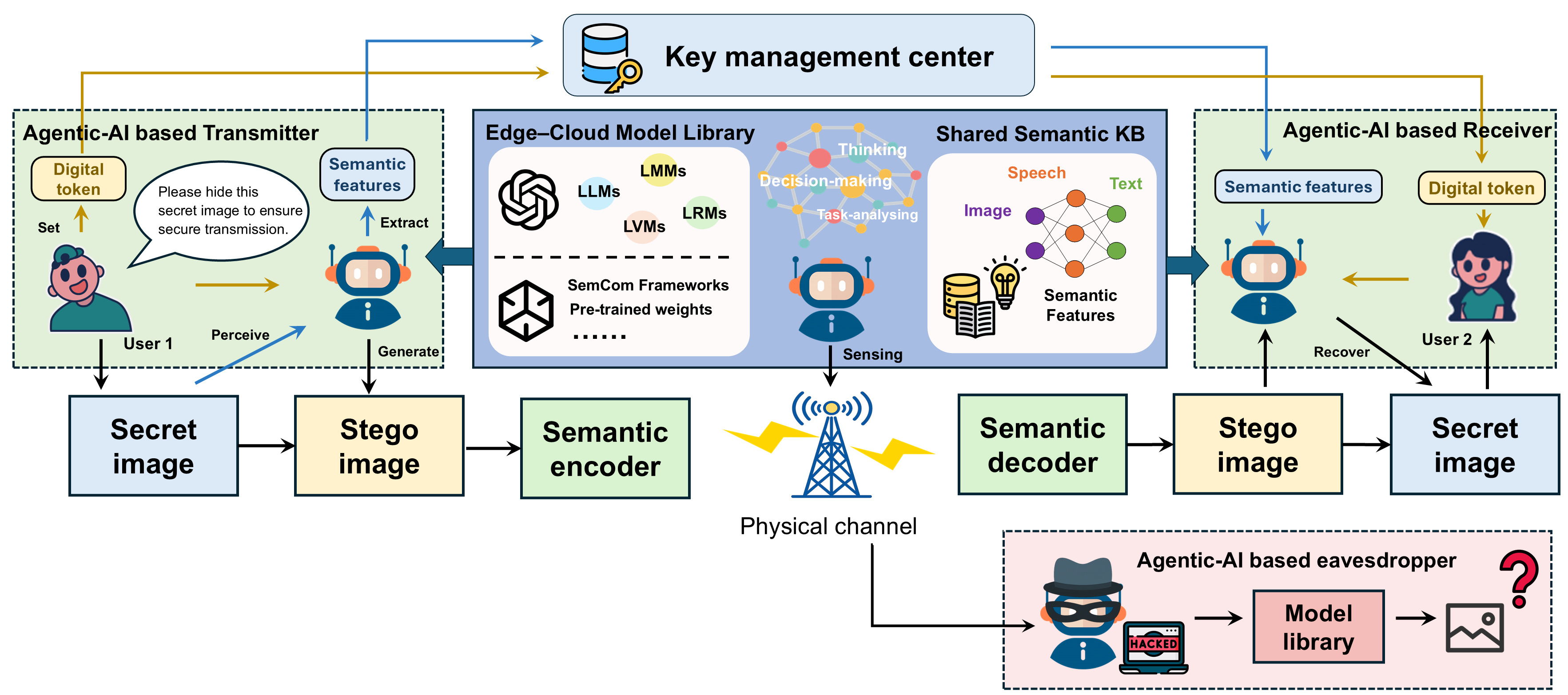}
\caption{Network model of the proposed AgentSemSteCom scheme, including the agentic AI-based transmitter, physical channel, agentic AI-based receiver, key management center, and agentic AI-based eavesdropper. It is realized based on the edge-cloud library \cite{zhang2024space-air-ground} and shared semantic knowledge base, with the aid of agentic AI's capabilities of environmental perception, task reasoning, and tool invocation. }
\label{framework}
\end{figure*}
As shown in Figure \ref{framework}, the proposed AgentSemSteCom scheme involves the following nodes, with main parameters illustrated in Table~\ref{tab:notations}.
\subsubsection{Agentic AI-based Transmitter}
Unlike traditional passive encoders, an intelligent agentic AI acts as the transmitter  with perception and tool-invocation capabilities \cite{gao2025agentic}. It autonomously analyzes the task requirements and content of the secret image $\mathbf{x}_s$ to dynamically select the optimal semantic feature extraction strategy. Agentic AI-based transmitter uses the features and user-defined digital token to hide secret images $\mathbf{x}_s$ into stego images $\mathbf{x}_{stego}$. 
Then, the transmitter performs environmental sensing to obtain the physical channel state, such as Signal-to-Noise Ratio (SNR) and available bandwidth \cite{jiang2025large}. The appropriate semantic model and knowledge base are shared between the legitimate transmitter and receiver. The transmitter encodes the stego image $\mathbf{x}_{stego}$ using semantic encoder and transmits it into channel.
\subsubsection{Physical Channel}
The wireless channel is modeled as a quasi-static fading
channel with Additive White Gaussian noise (AWGN) \cite{xu2023latent}. The noise component $n$ follows a complex Gaussian distribution $\mathcal{CN}(0, \sigma_w^2)$, where $\sigma_w^2$ indicates the noise power. Hence, the semantic vector of stego images transmitted through channel is
$\mathbf{\hat{\mathcal{S}}}_{stego} = h * \mathbf{\mathcal{S}}_{stego} + n$,
where $h$ is the coefficients of the physical channel.
\subsubsection{Agentic AI-based Receiver}
Agentic AI-based receiver first decodes the transmitted semantic vector into the reconstructed stego image $\mathbf{\hat{x}}_{stego}$. Then, it utilizes the semantic feature and digital token distributed by key management center to recover the secret image $\mathbf{\hat{x}}_s$. Depending on the autonomous task-oriented reasoning ability, it intelligently determines the subsequent action, such as denoising or super-resolution, to match the specific task requirements. 
\subsubsection{Key Management Center}

The key management center is responsible for user-defined digital token storage and distribution. Also, the shared semantic features and public keys between agentic AI are also managed by the center. It is assumed that the key management center provides a completely secure channel for successful transmission \cite{wang2025efficient}. To be noticed, the digital token is private while the public key and the semantic feature could be public.
\subsubsection{Agentic AI-based Eavesdropper}
The agentic AI-based eavesdropper attempts to intercept transmitted semantic information. It is assumed to have the access to semantic decoder and other available pre-trained Large AI Models (LAMs)\cite{zhang2024large}, the eavesdropper aims to recover secret images from steganographic semantic representation.

\subsection{Overview of the Proposed AgentSemSteCom scheme}

\begin{algorithm}[t]
\caption{The Steps of the proposed AgentSemSteCom.}
\label{alg:AgentSemSteCom}
\setlength{\baselineskip}{\baselineskip}
\begin{algorithmic}[1]
\Statex \noindent\hspace*{-1.8em} \textbf{Stage 1: Semantic Extraction and Public Key Generation}
\State Generate public key ${K}_{\text{pub}}$ from secret image $\mathbf{x}_s$  by ${K}_{\text{pub}} = f_\text{LLM}(\mathbf{x}_{s})$
\State Extract implicit feature $\mathbf{x}_{feat}$ via agentic decision function under task requirement by $\mathbf{x}_{feat} = \mathcal{G}(\mathbf{x}_\text{s},\mathcal{R})$

\Statex \noindent\hspace*{-1.8em} \textbf{Stage 2: Digital Token Controlled Reference Image Generation}
\State Initialize Gaussian noise $\mathbf{z}_D$ controlled by digital token $t$
\State Sample reference latent $\mathbf{z}_{ref}$ using DDIM under ${K}_{\text{pub}}$ and $\mathbf{x}_{feat}$ guidance by ${\mathbf{z}_{ref}} = \text{DDIM}(\mathbf{z}_D, \epsilon_\theta, K_{\text{pub}}, \mathbf{x}_{feat},T, 0)$
\State Decode $\mathbf{z}_{ref}$ to generate reference image $\mathbf{x}_{ref}$ via VAE decoder by $\mathbf{x}_{ref} = \mathcal{D}_\text{VAE}({\mathbf{z}}_{ref})$

\Statex \noindent\hspace*{-1.8em} \textbf{Stage 3: Coverless Stego Image Generation at Transmitter}
\State Encode secret image $\mathbf{x}_s$ into latent space $\mathbf{z}_s$ using VAE encoder by $\mathbf{z}_{s} = \mathcal{E}_\text{VAE}({\mathbf{x}}_{s})$
\State Perform forward diffusion to obtain noisy latent $\mathbf{z}_T$ using EDICT by $\mathbf{z}_T = \text{EDICT}(\mathbf{z}_s, \epsilon_\theta, 0, T)$
\State Apply deterministic latent noise perturbation to obtain $\mathbf{z}'_T$ by $\mathbf{z}'_T = f_\text{n}(\mathbf{z}_T)$
\State Perform reverse denoising under ${K}_{\text{pub}}$ and $\mathbf{x}_{ref}$ guidance to obtain stego latent $\mathbf{z}_{stego}$ by ${\mathbf{z}}_{stego} = \text{EDICT}(\mathbf{z}’_T, \epsilon_\theta, K_{\text{pub}},\mathbf{x}_{ref}, T, 0)$
\State Decode $\mathbf{z}_{stego}$ using VAE decoder to generate stego image $\mathbf{x}_{stego}$ by $\mathbf{x}_{stego} = \mathcal{D}_\text{VAE}({\mathbf{z}}_{stego})$

\Statex \noindent\hspace*{-1.8em} \textbf{Stage 4: JSCC-based Semantic Transmission}
\State Encode stego image $\mathbf{x}_{stego}$ into semantic feature $\mathcal{S}$ by semantic encoder using $\mathcal{S} = \mathcal{E}_{sem}(\mathbf{x}_{\text{stego}})$
\State Transmit $\mathcal{S}$ through AWGN channel to obtain $\mathcal{S}'$
\State Decode $\mathcal{S}'$ to reconstruct stego image $\hat{\mathbf{x}}_{stego}$ by semantic decoder using $\hat{\mathbf{x}}_{\text{stego}} =\mathcal{D}_{sem}(\mathcal{S}')$

\Statex \noindent\hspace*{-1.8em} \textbf{Stage 5: Secret Image Recovery at Receiver}
\State Encode received stego image $\hat{\mathbf{x}}_{stego}$ into latent representation $\hat{\mathbf{z}}_{stego}$
\State Perform forward diffusion under ${K}_{\text{pub}}$ and $\mathbf{x}_{ref}$ to obtain perturbed noisy latent $\hat{\mathbf{z}}'_T$ by $\hat{\mathbf{z}}_T' = \text{EDICT}(\hat{\mathbf{z}}_{stego}, \epsilon_\theta, K_{\text{pub}},\mathbf{x}_{ref}, 0, T)$
\State Restore noisy latent $\hat{\mathbf{z}}_T$ via inverse perturbation by $\hat{\mathbf{z}}_T = f_\text{n}^{-1}(\hat{\mathbf{z}}'_T)$
\State Perform reverse denoising to recover secret latent $\hat{\mathbf{z}}_0$ by $\hat{\mathbf{z}}_0 = \text{EDICT}(\hat{\mathbf{z}}_T, \epsilon_\theta, T, 0)$
\State Decode $\hat{\mathbf{z}}_0$ to obtain reconstructed secret image $\hat{\mathbf{x}}_s$

\Statex \noindent\hspace*{-1.8em} \textbf{Stage 6: Optional Task-oriented Enhancement}
\State Invoke optional task-specific enhancement modules to obtain enhanced image $\hat{\mathbf{x}}_s^{*}$ by $\hat{\mathbf{x}}_s^{*}=\mathcal{H}\!\left(
\hat{\mathbf{x}}_s \mid \mathcal{R}\right)$

\end{algorithmic}
\end{algorithm}

\begin{figure*}[]
\centering
\includegraphics[width=\linewidth]{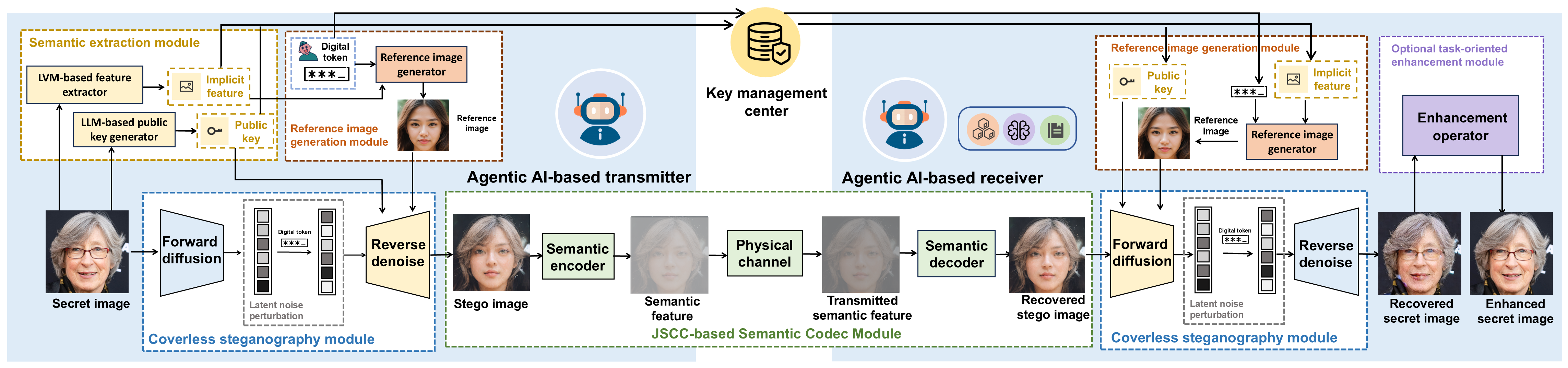}
\caption{The proposed AgentSemSteCom scheme, where the designed five modules include the semantic extraction module, digital token controlled reference image generation module, conditional diffusion model-based coverless steganography module, JSCC-based semantic codec module, and optional task-oriented enhancement module.}
\label{scheme}
\end{figure*}

\begin{table}[htbp]
\centering
\caption{The list of main parameters.}
\begin{tabular}{ll}
\toprule
\textbf{Notation} & \textbf{Meaning} \\
\midrule
$\alpha_t, \beta_t$ & Noise scheduling coefficients \\
$\mathcal{D}(\cdot)$ & Decoder \\
$\mathcal{E}(\cdot)$ & Encoder \\
$\epsilon$ & Random Gaussian noise \\
$\epsilon_\theta$ & Predicted noise component \\
$\eta$ & Perturbation strength coefficient \\
$k$ & Digital token \\
$K_{\text{pub}}$ & Public semantic key \\
$\mathbf{M}$ & Binary perturbation mask \\
$p$ & Mixing coefficient \\
$\mathcal{S}$ & Semantic feature \\
$s$ & Random seed \\
$T$ & Total diffusion steps \\
$t$ & Diffusion time step \\
$\mathbf{u}$ & Assistant latent variable in coupled sequences \\
$\mathbf{x}_{feat}$ & Implicit feature \\
$\mathbf{x}_{ref}$ & Reference image \\
$\mathbf{x}_s$ & Secret image \\
$\mathbf{x}_{stego}$ & Stego image \\
$\mathbf{z}$ & Primary latent variable during diffusion \\
\bottomrule
\end{tabular}
\label{tab:notations}
\end{table}

As presented in Figure \ref{scheme}, the proposed AgentSemSteCom scheme comprises five modules as follows.

\subsubsection{Semantic Extraction Module}
The semantic extraction module is used to extract public semantic key $K_{\text{pub}}$ and implicit feature $\mathbf{x}_{feat}$. Agentic AI invokes Large Language Models (LLMs) due to their stronger contextual understanding,
linguistic diversity, and control capability to generate the public key $K_{\text{pub}}$ as ${K}_{\text{pub}} = f_\text{LLM}({x}_{s})$, which determines the content of stego image $\mathbf{x}_{stego}$.
$K_{\text{pub}}= (K_1, K_2, \dots, K_T)$ is a sequence of semantic prompts. Based on the task requirement, agentic AI also adaptively constructs implicit feature according to the content characteristics of the secret image by $\mathbf{x}_{feat} = \mathcal{G}(x_\text{s},\mathcal{R})$,
where $\mathcal{G}(\cdot)$ is the dynamic decision function of the agent, and $\mathcal{R}$ is the task requirements autonomously perceived by agentic AI. 

\subsubsection{Digital Token Controlled Reference Image Generation Module}
This module is deployed at both transmitter and receiver side, and relies on a diffusion model based generator to synthesize reference image. The digital token controls the distribution of initial gaussian noise $\mathbf{z}_D$, which determines the start point of diffusion process. Also, to ensure natural semantic background and preserve the structure of secret image, the sampling of reference image is guided by $K_{\text{pub}}$ and implicit feature $\mathbf{x}_{feat}$ as ${\mathbf{z}_{ref}} = \text{DDIM}(\mathbf{z}_D, \epsilon_\theta, K_{\text{pub}}, \mathbf{x}_{feat},T, 0)$,
where $\epsilon_\theta$ is the noise prediction network. Then, the VAE-based decoder converts latent domain $\mathbf{z}_{ref}$ into bit domain to generate reference image $\mathbf{x}_{ref}$ as $\mathbf{x}_{ref} = \mathcal{D}_\text{VAE}({\mathbf{z}}_{ref})$.

\subsubsection{Conditional Diffusion Model-based Coverless Steganography Module}
To realize coverless stego image generation, the conditional diffusion model-based module directly generates stego images from latent representations, which is guided by public semantic key and reference image at sampling stages.
\paragraph{Transmitter}
At the transmitter, the secret image is first transformed into a latent vector $\mathbf{z}_s$ by a VAE-based encoder $\mathcal{E}(\cdot)$ as $\mathbf{z}_{s} = \mathcal{E}_\text{VAE}({\mathbf{x}}_{s})$.
Then, the noisy latent representation $\mathbf{z}_T$ is obtained in forward diffusion process by a deterministic sampling method as 
$\mathbf{z}_T = \text{EDICT}(\mathbf{z}_s, \epsilon_\theta, 0, T)$. Furthermore, the noisy latent vector $\mathbf{z}_T$ is added deterministic perturbation $f_\text{n}(\cdot)$ controlled by digital token, which serves as a random seed to generate a binary mask, flipping the signs of certain positions as $\mathbf{z}'_T = f_\text{n}(\mathbf{z}_T)$.
Subsequently, the public semantic key $K_{\text{pub}}$ and reference image $\mathbf{x}_{ref}$ are used to get the stego latent representation $\mathbf{z}_0$ by guiding the denoising process as ${\mathbf{z}}_{stego} = \text{EDICT}(\mathbf{z}’_T, \epsilon_\theta, K_{\text{pub}},\mathbf{x}_{ref}, T, 0)$,
Finally, the VAE-based decoder $\mathcal{D}(\cdot)$ converts latent domain $\mathbf{z}_0$ into bit domain to reconstruct stego image $\mathbf{x}_{\text{stego}}$ by $\mathbf{x}_{stego} = \mathcal{D}_\text{VAE}({\mathbf{z}}_{stego})$.

\paragraph{Receiver}
At the receiver, we follow the same procedure as at the transmitter, but exchange the order of forward diffusion and denoising to achieve the inverse process. After operating reverse denoising for latent representation of transmitted stego image $\hat{\mathbf{z}}_{stego}$ by $\hat{\mathbf{z}}_T' = \text{EDICT}(\hat{\mathbf{z}}_{stego}, \epsilon_\theta, K_{\text{pub}},\mathbf{x}_{ref}, 0, T)$, the recovery perturbed noisy latent vector $\hat{\mathbf{z}}'_T$ is restored back to $\hat{\mathbf{z}}_T$ by $\hat{\mathbf{z}}_T = f_\text{n}^{-1}(\hat{\mathbf{z}}'_T)$. Then, the recovery latent vector of secret image is obtained by reverse forward diffusion as $\hat{\mathbf{z}}_0 = \text{EDICT}(\hat{\mathbf{z}}_T, \epsilon_\theta, T, 0)$,



\subsubsection{Joint Source-Channel Coding (JSCC)-based Semantic Codec Module}
A typical SemCom transmission is considered over a wireless channel, where the goal is to transmit the stego image with minimal semantic distortion under noisy wireless channel. To this end, a JSCC-based semantic codec module is employed that directly learns a mapping between the visual domain and physical channel domain\cite{bourtsoulatze2019deep}. Specifically, the stego image $\mathbf{x}_{\text{stego}}$ is encoded into a semantic feature $\mathcal{S}$ through a semantic encoder $\mathcal{E}_{sem}(\cdot)$ by $\mathcal{S} = \mathcal{E}_{sem}(\mathbf{x}_{\text{stego}})$.
This latent representation is directly transmitted over the wireless channel, yielding a distorted signal $\mathcal{S}'$. At the receiver, a decoder $\mathcal{D}_{sem}(\cdot)$ is employed to reconstruct the image $\hat{\mathbf{x}}_{\text{stego}}$ by $\hat{\mathbf{x}}_{\text{stego}} =\mathcal{D}_{sem}(\mathcal{S}')$.

\subsubsection{Optional Task-oriented Enhancement Module}
To further enhance the adaptability of SemCom, we introduce an optional task-oriented enhancement module driven by agentic AI. 
By perceiving the reconstruction quality of the secret image $\hat{x}_s$ and evaluating the task requirement, agentic AI-based receiver can identify the most effective enhancement scheme tailored to current channel conditions and content types \cite{liu2025lameta} by $\hat{\mathbf{x}}_s^{*}=\mathcal{H}\!\left(
\hat{\mathbf{x}}_s \mid \mathcal{R}\right)$,
where $\mathcal{H}(\cdot)$ denotes a task-oriented enhancement operator selected by agentic AI. Depending on the scenario, it can invoke specialized tools from a shared model library, such as denoising models for low SNR environments or super-resolution models for high-fidelity tasks like facial recognition.

Overall, the general steps of AgentSemSteCom are summarized in Algorithm \ref{alg:AgentSemSteCom}.



\section{Public Semantic Key and Implicit Feature Extraction}
Both public semantic keys and implicit features are established as public conditions that contain no reversible or confidential information, enabling a private keyless approach. The public semantic key ensures the reasonable generation of stego image, while the implicit feature preserves structural  constraint to enable the semantic reconstruction in spatial organization. These conditions are analyzed by agentic AI to jointly guide the diffusion-based image generation process.

\subsection{LLM-driven Public Semantic Key Generation}
In this work, the public semantic key does not depend on the content of private semantic key. Instead, it is directly generated by agentic AI after perceptual analysis of the secret image. The purpose of the public semantic key is not to carry reversible information, but to provide a semantically plausible generation context that can be naturally integrated into the diffusion process. This design ensures that the generating stego image follows a statistically normal semantic distribution and does not raise suspicion from an eavesdropper. 
Specifically, agentic AI invokes a LLM as a generative tool to produce the public key sequence in an autoregressive manner \cite{zhang2025toward,liu2024hierarchical}. At the $\ell$-th layer, the representation of public prompts $\mathbf{q}_{\ell}$ is updated as 
\begin{equation}
\label{csa}
\mathbf{q}'_{\ell} = \mathrm{CSA}\big(\mathrm{LN}(\mathbf{q}_{\ell-1})\big) + \mathbf{q}_{\ell-1},
\end{equation}
\begin{equation}
\label{mlp}
\mathbf{q}_{\ell} = \mathrm{MLP}\big(\mathrm{LN}(\mathbf{q}'_{\ell})\big) + \mathbf{q}'_{\ell},
\end{equation}
where $\mathrm{CSA}(\cdot)$ denotes the causal self-attention operator, $\mathrm{LN}(\cdot)$ is the layer normalization, $\mathbf{q}'_{\ell}$ is the intermediate representation after attention, and $\mathrm{MLP}(\cdot)$ denotes the position-wise feed-forward network.
During decoding, the prompt distribution $\hat{\mathbf{K}}_{i}$ is computed from the final hidden representation \cite{li2022blip} at the last layer $L$ as 
\begin{equation}
\label{soft}
\hat{\mathbf{K}}_{i} = \mathrm{Softmax}\big(\mathbf{W}_{\text{pub}} \cdot \mathrm{LN}(\mathbf{q}_{L}^{i})\big), \quad i=1,\ldots,T,
\end{equation}
where $\hat{\mathbf{K}}_{i}$ is the probability vector over the public vocabulary at position $i$, $\mathbf{W}_{\text{pub}}$ is the output projection matrix that maps hidden states to vocabulary logits, and $\mathrm{Softmax}(\cdot)$ is applied to the vocabulary dimension. 
From $\hat{\mathbf{K}}_{i}$, $K_i$ is sampled at each position, yielding the final generated public semantic key $K_{\text{pub}} = (K_1, K_2, \ldots, K_T)$.

\subsection{Adaptive Implicit Feature Extraction}
Semantic-text-key-only guidance often results in generated images whose structural layout deviates noticeably from that of the secret image, which may increase the suspicion of eavesdroppers \cite{yang2024diffstega}. To address this issue, the implicit feature $\mathbf{x}_{feat}$is introduced as an extra constraint. Agentic AI relies on its perception ability to extract the deep semantic feature, which reflects the underlying spatial distribution of the secret image. The implicit feature not only guides the generation of the reference image to remain structurally aligned with the secret image, thereby preventing excessive visual divergence in the generated stega image, but also preserves critical structural cues for accurate reconstruction of the secret image at the receiver. Note that the implicit feature is set public, since the extracted structure carries no sensitive semantic information.



For the common scenario, agentic AI extracts semantic segmentation map as the implicit feature, such as the general background and main object.
When agentic AI detects that the secret image involves high-fidelity scenarios such as face detection, the implicit feature extraction process is further refined toward deeper structural abstraction, progressing from pose perception to fine-grained skeleton reconstruction. In this setting, agentic AI leverages the accurate modeling of local geometric configurations and inter-keypoint relationships to extract the implicit feature $\mathbf{x}_{feat}$, which is a skeleton graph jointly formed by connected edges \cite{cao2017realtime}.

\section{Digital Token Controlled Reference Image Generation}
The reference image embeds the implicit semantic feature into specific visual content to provide a controllable condition for the stego image generation, which is obtained by a reference generator. The agentic AI considers two steps to obtain the controllable reference images, including digital token initialized latent noise and the ControlNet-based diffusion generation.
\subsection{Digital Token Initialized Latent Noise}
Motivated by \cite{yang2024diffstega}, we introduce digital token, a user-defined numerical sequence with variable length, to deterministically initialize the reference image generation. Each digital token leads to one specific reference image. This mapping relationship ensures that only the legitimate receiver with correct digital token can generate relative reference image, and recover the original secret image. By selecting different digital tokens, users can encrypt the secret image into various stego images without introducing any private semantic key or other sensitive information.

As agentic AI receives the digital token $k$, it converts the numbers into a random seed through the hash function $\mathcal{H}(\cdot)$ by 
\begin{equation}
\label{seed}
s = \mathcal{H}(k).
\end{equation}
Using this seed, agentic AI samples the initial latent noise $\mathbf{z}_D$ from a standard Gaussian distribution as $\mathbf{z}_D \sim \mathcal{N}(\mathbf{0}, \mathbf{I})$, 
where $\mathbf{z}_D = \mathrm{Randn}(s)$, and $\mathrm{Randn}(s)$ represents Gaussian noise sampling under seed $s$.
By fixing the start point of the diffusion process based on the digital token, agentic AI ensures the entire generation trajectory is deterministic and reproducible. Given the same token, both the legitimate transmitter and receiver can reconstruct an identical latent initialization without exchanging any additional private information.

\subsection{ControlNet-based Diffusion for Reference Image Generation}
\begin{algorithm}[t]
\caption{ControlNet-based Diffusion for Reference Image Generation}
\label{alg:controlnet_ref}
\setlength{\baselineskip}{\baselineskip}
\begin{algorithmic}[1]
\Statex \hspace{-\algorithmicindent} \textbf{Training Phase:}
\Repeat
    \State Sample a training image $\mathbf{x}_0 \sim q(\mathbf{x})$ and its paired implicit feature $\mathbf{x}_{feat}$
    \State Compute text-condition embeddings $\mathbf{c}_t = f_{\text{text}}(K_{\text{pub}})$
    \State Compute structural-condition embeddings 
    \State \hspace{1.2em} $\mathbf{c}_f = f_{\text{feat}}(\mathbf{x}_{feat})$
    \State Obtain the latent vector $\mathbf{z}_0$ by $\mathbf{z}_{s} = \mathcal{E}_\text{VAE}({\mathbf{x}}_{s})$
    \State Sample diffusion step $t \sim \mathcal{U}\{1,\ldots,T\}$ and noise $\epsilon \sim \mathcal{N}(0,\mathbf{I})$
    \State Construct the noisy latent variable 
    \State \hspace{1.2em} $\mathbf{z}_t = \sqrt{\alpha_t}\mathbf{z}_0 + \sqrt{1-\alpha_t}\epsilon$
    \State Compute the controlled block output $y$ via zero-convolution injection by (\ref{eq:controlnet_inject})
    \State Predict noise $\hat{\epsilon}$ using the ControlNet backbone by (\ref{feature_noise})
    \State Compute the denoising objective $\mathcal{L}$ by (\ref{eq:controlnet_loss})
    \State Update $\Theta_c, \Theta_{z_1}, \Theta_{z_2}$  via gradient descent while keeping $\Theta$ frozen
\Until{converged}
\vspace{0.1em}
\Statex \hspace{-\algorithmicindent} \textbf{Inference Phase }
\Statex \textbf{Input:} $\mathbf{x}_{feat}$, $K_{\text{pub}}$ and digital token controlled noise $\mathbf{z}_D$
\State Initialize the sampling latent $\mathbf{z}_T = \mathbf{z}_D$
\For{$t=T,T-1,\ldots,1$}
    \State Predict noise by (\ref{feature_noise})
    \State Update latent by DDIM sampling by (\ref{reference_sampling})
\EndFor
\State Decode the final latent to obtain the reference image $\mathbf{x}_{ref}$ by $\mathbf{x}_{ref} = \mathcal{D}(\mathbf{z}_0)$
\State \Return reference image $\mathbf{x}_{ref}$
\end{algorithmic}
\end{algorithm}

To ensure that the reference image $\mathbf{x}_{ref}$ strictly follows the structural constraints provided by the implicit feature $\mathbf{x}_{feat}$, agentic AI invokes ControlNet as a structure-conditioning injector in the reference image generator. It allows the reference image to retain the structure of secret image while the visual pattern is determined by the public semantic key $K_{\text{pub}}$.

The ControlNet is implemented as a trainable side-branch attached to the frozen U-Net backbone of the diffusion model \cite{zhang2023adding}. To preserve the pretrained generative priors, the parameters $\Theta$ of the original U-Net remain fixed, while agentic AI clones all encoding blocks and middle blocks into a parallel trainable branch with parameters $\Theta_c$. This branch is dedicated to extracting  structural representations from the implicit feature $\mathbf{x}_{feat}$, which are progressively injected into the frozen U-Net through zero-initialized convolution layers $\mathcal{Z}(\cdot;\Theta_z)$.
The controlled output $y$ of a U-Net block at diffusion step $t$ is expressed as
\begin{equation}
\label{eq:controlnet_inject}
y = \mathcal{F}(\mathbf{z}_t; \Theta) + \mathcal{Z}\Big(\mathcal{F}\big(\mathbf{z}_t + \mathcal{Z}(c; \Theta_{z_1}); \Theta_c\big); \Theta_{z_2}\Big),
\end{equation}
where $\mathcal{F}(\cdot;\Theta)$ denotes the $K_{\text{pub}}$ guided transformation of a specific block in the frozen U-Net, and $\mathcal{F}(\cdot;\Theta_c)$ represents the trainable $\mathbf{x}_{feat}$ guided transformation in the ControlNet branch. The zero-convolution layers, parameterized by $\Theta_{z_1}$ and $\Theta_{z_2}$, are initialized to zero, ensuring that conditions are gradually introduced during training, thereby maintaining generative stability.

Although the intermediate feature $y$ is not directly supervised, it participates in the subsequent U-Net forward propagation and ultimately influences the noise prediction $\epsilon_\theta(\cdot)$. Consequently, the parameters of ControlNet are optimized implicitly through the standard diffusion objective:
\begin{equation}
\begin{aligned}
\label{eq:controlnet_loss}
\mathcal{L} = \mathbb{E}_{\mathbf{z}_0, t, \mathbf{x}_{feat},K_{\text{pub}}, \epsilon \sim \mathcal{N}(0,1)} \Big[ | \epsilon - \epsilon_\theta(\mathbf{z}_t, t, \mathbf{x}_{feat}, K_{\text{pub}}) |_2^2 \Big], 
\end{aligned}
\end{equation}
where $\epsilon_\theta$ is the noise predictor modified by the ControlNet branch. Through back propagation, the trainable parameters $\Theta_c$, $\Theta_{z_1}$, and $\Theta_{z_2}$ are updated, while the backbone parameters $\Theta$ remain fixed.
At the inference stage, starting from the token initialized latent variable $\mathbf{z}_D$, agentic AI iteratively samples the reference image using the Denoising Diffusion Implicit Models (DDIM) strategy \cite{song2020denoising} from $t$ to $t-1$ as
\begin{equation}
\label{reference_sampling}
\mathbf{z}_{t-1} = \sqrt{\alpha_{t-1}}  \frac{\mathbf{z}_t - \sqrt{1-\alpha_t} \hat{\epsilon}_\theta}{\sqrt{\alpha_t}} + \sqrt{1-\alpha_{t-1}} \hat{\epsilon}_\theta,
\end{equation}
\begin{equation}
\label{feature_noise}
\hat{\epsilon}_\theta = \epsilon_\theta(\mathbf{z}_t, t, \mathbf{x}_{feat}, K_{\text{pub}}),
\end{equation}
where $\alpha$ is the noise scheduling coefficient. Finally, agentic AI decodes $\mathbf{z}_0$ through the VAE decoder $\mathcal{D}(\cdot)$ to obtain the reference image $\mathbf{x}_{ref} = \mathcal{D}(\mathbf{z}_0)$. The whole process is shown in \textbf{Algorithm \ref{alg:controlnet_ref}}.


\section{Conditional Diffusion Model-based Coverless Steganography}
\begin{figure*}[]
\centering
\includegraphics[width=0.9\linewidth]{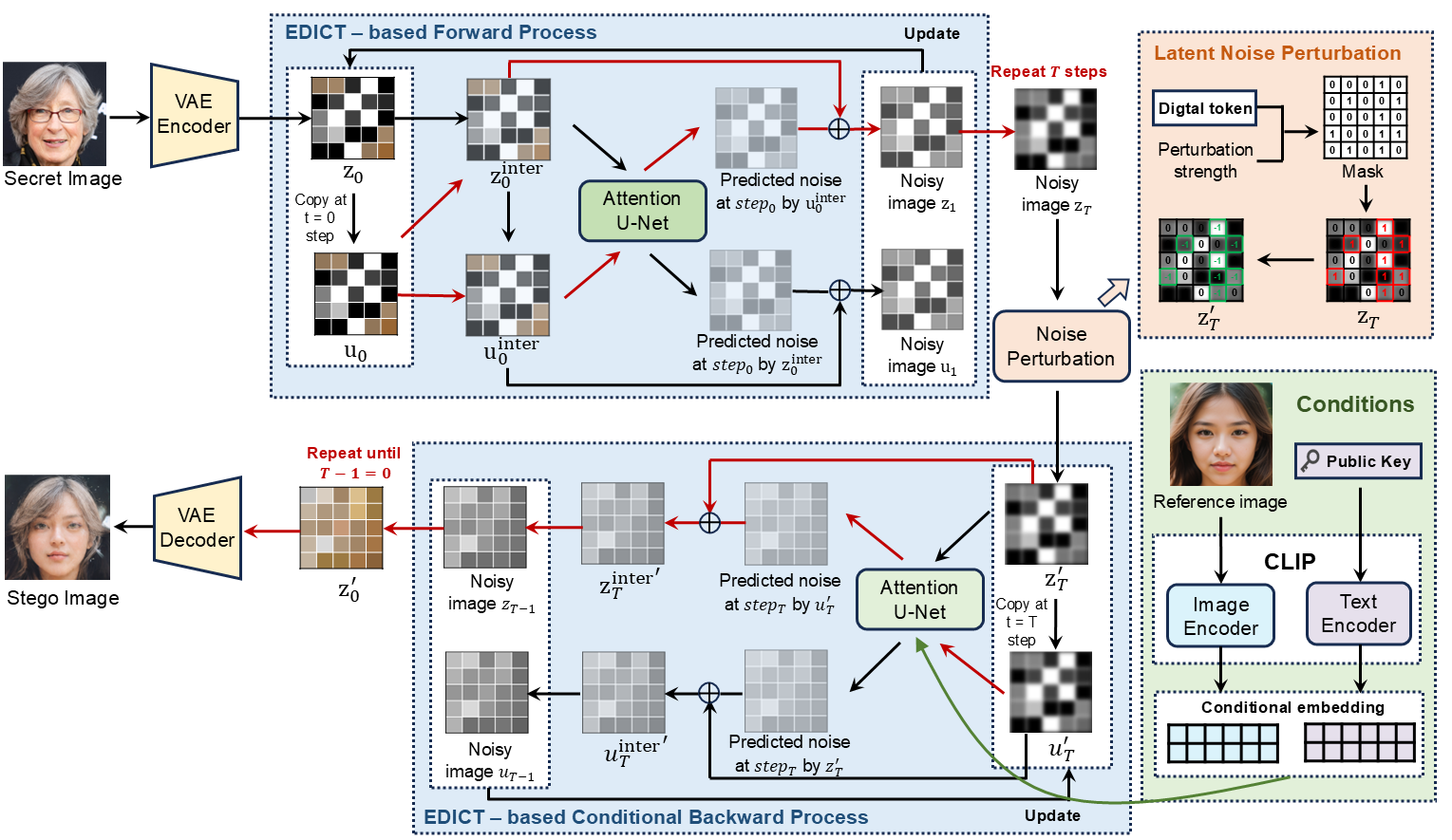}
\caption{The diagram for conditional diffusion model-based coverless steganography module, where the sampling is based on EDICT and the diffusion process is guided by reference images and public text. The noise perturbation is achieved by flipping specific coordinates of latent vector, which are determined by digital token and perturbation strength.}
\label{fig：edict}
\end{figure*}

The conditional diffusion model-based coverless steganography module is employed to realize secure embedding of secret images into generated stego images without relying on traditional cover carriers. To achieve this objective, the agentic AI adopts an EDICT-based exact invertible process to eliminate sampling errors, a digital token-driven latent perturbation to enhanced security, and a decoupled cross-attention mechanism to precisely control the generative trajectory. The detailed diagram is shown in Figure \ref{fig：edict}.

\subsection{EDICT-based Invertible Sampling and Diffusion Process}

To ensure high-fidelity transmission of secret semantics in complex channel environments, the Exact Diffusion Inversion via Coupled Transformations (EDICT) \cite{wallace2023edict} is incorporated into the generative logic of agentic AI, which mitigates sampling errors caused by linearization approximation during DDIM sampling. Unlike traditional single-chain paths, agentic AI maintains a pair of coupled latent variable sequences $(\mathbf{z}_t, \mathbf{u}_t)$. By leveraging the reversibility of affine coupling layers, this dual-track design shown in  (\ref{edict_z}) and (\ref{edict_u}) mathematically enforces exact alignment between the forward and reverse processes.
\begin{equation}
\label{edict_z}
    z_{t-1} = \alpha_t \mathbf{z}_t + \beta_t \epsilon_\theta(\mathbf{u}_t, t, c),
\end{equation}
\begin{equation}
\label{edict_u}
    u_{t-1} = \alpha_t \mathbf{u}_t + \beta_t \epsilon_\theta(z_{t-1}, t, c),
\end{equation}
where $\epsilon_\theta$ denotes the noise component predicted by the agent based on current state and conditions $c$. $\alpha_t$ and $\beta_t$ are coefficients derived from the noise schedule. 
The state $\mathbf{u}_t$ is used to assist the update of $\mathbf{z}_t$, while the updated intermediate state $z_{t-1}$ serves as the basis to update $\mathbf{u}_t$ in an alternating fashion. This design ensures that the complex non-linear noise prediction term $\epsilon_\theta$ relies solely on the coupled counterpart rather than the variable being updated. Consequently, this transformation remains affine with respect to the target variable, allowing agentic AI to perform mathematically exact inversion by simple calculation, which fundamentally eliminates the linearization errors in diffusion processes.

To prevent the two coupled latent sequences from evolving independently, agentic AI strategically introduces a mixing layer between the coupling updates. This layer acts as a synchronization mechanism, forcing an information interaction between the primary and assistant chains. Agentic AI executes this mixing operation using a linear weighted combination, defined as $\mathbf{z}_t' = p \mathbf{z}_t + (1 - p) \mathbf{u}_t$, 
where $p$ is a mixing coefficient between 0 and 1. By regulating this parameter, agentic AI balances the independence of chains with the necessary mutual information flow. To maintain the stability of the system, agentic AI dynamically adjusts the coefficient $p$ according to the transmission environment and specific scenarios. 

Based on this reversible EDICT sampling method, agentic AI achieves the steganographic process through two symmetric phases. By combining the forward deterministic noising process and sampling denoising diffusion, agentic AI guarantees that the trajectory from the secret image to the noise representation is entirely deterministic and lossless.
\paragraph{Forward deterministic noising} In the forward stage, Agentic AI performs deterministic noising to map the secret image coupled state $(z_0, u_0)$ to the latent noise $(\mathbf{z}_t, \mathbf{u}_t)$ over $T$ steps. As described in (\ref{edict_forward}),  agentic AI first calculates the mixed states to obtain the intermediate variables $(z^{inter}_{t+1}, u^{inter}_{t+1})$ by inverting the mixing layer, and then reverses the affine coupling transformation to compute the final states $(z_{t+1}, u_{t+1})$.
\begin{subequations}
\label{edict_forward}
\renewcommand{\theequation}{\theparentequation-\arabic{equation}}
\begin{align}
u^{inter}_{t+1} &= (\mathbf{u}_t - (1 - p) \cdot \mathbf{z}_t)/p,  \label{edict_forward_a}\\
z^{inter}_{t+1} &= (\mathbf{z}_t - (1 - p) \cdot u^{inter}_{t+1})/p, \label{edict_forward_b}\\
u_{t+1} &= \gamma_{t+1}u^{inter}_{t+1} - \omega_{t+1} \cdot \epsilon_\theta(z^{inter}_{t+1}, t+1), \label{edict_forward_c}\\
z_{t+1} &= \gamma_{t+1}z^{inter}_{t+1} - \omega_{t+1} \cdot \epsilon_\theta(u_{t+1}, t+1),\label{edict_forward_d}
\end{align}
\end{subequations}
where $\gamma_t = \alpha_t^{-1}$ and $\omega_t = \beta_t \alpha_t^{-1}$.
\vspace{0.5em}
\paragraph{Sampling Denoising diffusion}In the backward stage, agentic AI executes the sampling denoising diffusion to generate the stego latent vector from the perturbed noise vector $z'_T$ , which inversely applies the $T$ steps of forward noising as follow:
\begin{subequations}
\label{edict_reverse}
\renewcommand{\theequation}{\theparentequation-\arabic{equation}}
\begin{align}
z'^{inter}_t &= \alpha_t \cdot z'_t + \beta_t \cdot \epsilon_\theta(u'_t, t), \label{edict_reverse_a}\\
u'^{inter}_t &= \alpha_t \cdot u'_t + \beta_t \cdot \epsilon_\theta(z'^{inter}_t, t), \label{edict_reverse_b}\\
z'_{t-1} &= p \cdot z'^{inter}_t + (1 - p) \cdot u'^{inter}_t, \label{edict_reverse_c}\\
u'_{t-1} &= p \cdot u'^{inter}_t + (1 - p) \cdot z'_{t-1}.\label{edict_reverse_d}
\end{align}
\end{subequations}

\subsection{Deterministic Latent Noise Perturbation}


Following the execution of the deterministic forward noising process in  (\ref{edict_forward}), the agentic AI-based transmitter obtains the latent noise $\mathbf{z}_T$ of the secret image $\mathbf{x}_s$. To strengthen the security of the proposed method, the digital token also serves as a password to perturb the latent state \cite{yang2024diffstega}. This design reuses the user-defined token and only needs to transmit it once, which effectively eliminates the risk of information leakage associated with frequent secret key exchanges. Utilizing the seed in (\ref{seed}), agentic AI constructs a binary perturbation mask $\mathbf{M} \in \{0, 1\}^{C \times H \times W}$ through a pseudo-random generation process. 
\paragraph{Digital Token-guided Latent Noise Inversion}
At the transmitter side, the mask is added to $\mathbf{z}_T$ to obtain $\mathbf{z}'_T$ by
\begin{equation} 
\label{perturbation} 
\mathbf{z}'_{T} = \mathbf{z}_{T} \odot (1 - \mathbf{M}(s, \eta)) - \mathbf{z}_{T} \odot \mathbf{M}(s, \eta), \end{equation}
where $\eta$ is the perturbation strength coefficient. A large coefficient leads to a strong perturbation. 
Since the agent employs a standard Gaussian distribution during the initial noising phase, this inversion operation preserves the statistical properties of the latent space.

\paragraph{Symmetric Restoration of Recovery Stego Vector}
At the receiver side, agentic AI performs a symmetric restoration of the latent state. After recovering the stego latent vector $\hat{z}'_T$ guided by semantic features, the agentic AI-based receiver uses the same digital token distributed by the key management center to calculate the mask $\mathbf{M}$. The restoration process is formulated as:
\begin{equation}
\label{restoration}
 \hat{\mathbf{z}}_{T} = \hat{\mathbf{z}}'_{T} \odot (1 - \mathbf{M}(s, \eta)) - \hat{\mathbf{z}}'_{T} \odot \mathbf{M}(s, \eta). 
 \end{equation}
The symmetric inverse process ensures the legitimate receiver can perfectly restore $\hat{z}_T$. 

\subsection{Public Semantic Key and Reference Image Guided Stego Image Generation}
To ensure that the accurate and expected secret image can be reconstructed from the stego image,  agentic AI adapts the decoupled cross-attention mechanism into the framework. 
Specifically, agentic AI initiates an encoding process to prepare the multimodal inputs. For the textual guidance, the public semantic key is processed by the text encoder of CLIP $f_\text{text}(\cdot)$ \cite{radford2021learning} to obtain the text embeddings $p_{text}$ as
\begin{equation}
\label{EncoderText}
p_{text} = f_\text{text}(\mathbf{K}_{pub}).
\end{equation}
For the visual guidance, agentic AI leverages an image encoder to extract the high-fidelity global image feature $v$ from the reference image $\mathbf{x}_{ref}$ as
\begin{equation}
\label{Encoderimage}
v = f_\text{image}(\mathbf{x}_{ref}),
\end{equation}
To effectively translate this static global embedding into a dynamic sequence that the diffusion model can utilize, agentic AI employs a trainable projection network. It can ensure the dimension of image feature is aligned with that of the text features \cite{ye2023ip}. It consists of a linear projection layer followed by layer normalization \cite{ba2016layer}. The processed image embeddings through this transformation are expressed as
\begin{equation} 
\label{projection}
p_{img} = \text{LN}(\mathbf{W}_{proj}v + \mathbf{b}_{proj}), 
\end{equation}
where $\mathbf{W}_{proj}$ and $\mathbf{b}_{proj}$ represent the weight and bias of the linear layer, respectively. 

Subsequently, agentic AI utilizes the decoupled cross-attention mechanism to integrate these multimodal features into the U-Net \cite{ronneberger2015u} intermediate layers. Unlike traditional cross-attention strategies \cite{cheng2026apeg}, agentic AI independently computes the attention maps for text features and image features in separate attention layers to prevent interference, rather than competing in a single shared space. The text guided attention \cite{vaswani2017attention} is shown as
\begin{equation}
\label{textAtten}
\begin{aligned}
{Z}_{text} 
&= \text{Attention}({Q_{text}}, {K_{text}}, {V_{text}}) \\
&= \text{Softmax}\left(\frac{{Q_{text}}{K_{text}}^\top}{\sqrt{d}}\right){V_{text}}.
\end{aligned}
\end{equation}
The keys, values, and query used above are obtained by projecting text feature $p_{text}$ as
\begin{equation}
\label{QKV1}
Q_{text} = \mathbf{z}_tW_Q, \quad K_{text} = p_{text}W_K, \quad V_{text} = p_{text}W_V,
\end{equation}
where $W_Q$, $W_K$, and $W_V$ are trainable parameters, and $t$ is the sampling step. 
The image guided attention counterpart is as follows:
\begin{equation}
\label{imageAtten}
\begin{aligned}
{Z}_{img} 
&= \text{Attention}({Q_{img}}, {K_{img}}, {V_{img}}) \\
&= \text{Softmax}\left(\frac{{Q_{img}}{K_{img}}^\top}{\sqrt{d}}\right){V_{img}},
\end{aligned}
\end{equation}
\begin{equation}
\label{QKV2}
Q_{img} = \mathbf{z}_tW_Q , \quad K_{img} = p_{img}W_K , \quad V_{img} = p_{img}W_V.
\end{equation}

Finally, agentic AI decouples these two distinct flows as conditions for the diffusion model. The conditional latent vector $\mathbf{Z}_{con}$ fed into the subsequent layer is defined as
\begin{equation}
\begin{aligned}
Z_{\text{con}}
&= Z_{\text{text}} + Z_{\text{img}} \\
&= \mathrm{Softmax}\!\left(
\frac{Q_{\text{text}} K_{\text{text}}^\top}{\sqrt{d}}
\right) V_{\text{text}} \\
&\quad + \mathrm{Softmax}\!\left(
\frac{Q_{\text{img}} K_{\text{img}}^\top}{\sqrt{d}}
\right) V_{\text{img}} .
\end{aligned}
\label{DecoupledAttention}
\end{equation}

\begin{algorithm}[t]
\caption{Public Semantic
Key and Reference Image Guided Stego Image Generation}
\label{alg:ipadapter_edict}
\setlength{\baselineskip}{\baselineskip}
\begin{algorithmic}[1]
\Statex \hspace{-\algorithmicindent} \textbf{Training Phase :}
\Repeat
    \State Sample training image $\mathbf{x}_0 \sim q(\mathbf{x})$
    \State Encode $K_{\text{pub}}$ and $\mathbf{x}_{ref}$ to obtain embeddings $p_{text}$ and $p_{img}$ by (\ref{EncoderText}) -- (\ref{projection})
    \State Sample diffusion step $t \sim \mathcal{U}\{1,\dots,T\}$ and noise $\epsilon \sim \mathcal{N}(0,\mathbf{I})$
    \State Encode image $\mathbf{x}_0$ and construct noisy latent $\mathbf{z}_t$
    \State Compute the attention feature $Z_{text}$ and $Z_{img}$  by (\ref{textAtten})--(\ref{QKV2})
    \State Obtain decoupled cross-attention
 $Z_{con}$ by (\ref{DecoupledAttention})
    \State Predict image-text pairs noise $\hat{\epsilon}_\theta(\mathbf{z}_t,t,K_{\text{pub}},\mathbf{x}_{ref})$
    \State Compute the loss function $\mathcal{L} = \|\epsilon - \hat{\epsilon}_\theta\|_2^2$
    \State Update the projection network parameters while freezing diffusion backbone
\Until{converged}
\vspace{0.1em}
\Statex \hspace{-\algorithmicindent} \textbf{Inference Phase }
\Statex \textbf{Input:} $\mathbf{x}_s$, $K_{\text{pub}}$, $\mathbf{x}_{ref}$
\State Encode secret image $\mathbf{x}_s$ into latent representation $\mathbf{z}_0$
\State Initialize coupled latent states $(\mathbf{z}_0,\mathbf{u}_0)$
\For{$t=0$ \textbf{to} $T-1$}
    \State Compute the noise latent vector $(\mathbf{z}_t,\mathbf{u}_t)$ using forward EDICT process by (\ref{edict_forward})
\EndFor
\For{$t=T$ \textbf{to} $1$}
    \State Predict $K_{\text{pub}}$-only noise $\epsilon_\theta(\mathbf{u}_t,t,K_{\text{pub}})$ 
    \State Predict $\mathbf{x}_{ref}$-conditioned noise $\epsilon_\theta(\mathbf{u}_t,t,K_{\text{pub}},\mathbf{x}_{ref})$ using the trained projection network:
    \State Compute conditional noise of $\mathbf{u}_t$ with image guidance weight $\lambda$ \cite{ho2022classifier}:
    \State \hspace{1.2em} 
    \begin{equation*}
    \begin{aligned}
    \tilde{\epsilon}^{(u)}_{t,c} 
    &= (1-\lambda)\,\epsilon_\theta(\mathbf{u}_t,t,K_{\text{pub}})\\
    &\quad + \lambda\,\epsilon_\theta(\mathbf{u}_t,t,K_{\text{pub}},\mathbf{x}_{ref})
    \end{aligned}
    \end{equation*}
    
    \State Compute the intermediate state $\mathbf{z}_t^{\mathrm{inter}}$ by (\ref{eq:edict_rev_inter}) 
    \State Compute conditional noise $\tilde{\epsilon}^{(z)}_{t,c}$ of $\mathbf{z}_t$ on the coupled chain with the same steps as $\tilde{\epsilon}^{(u)}_{t,c}$
    \State Compute the final conditional noise in $t$ step by (\ref{eq:edict_rev_fused_eps}):
    \State Update sampling result $z_{t-1}$ by (\ref{eq:edict_rev_ddimlike})
\EndFor
\State Decode $\mathbf{z}_{stego}$ to obtain stego image $\mathbf{x}_{stego}$
\State \Return stego image $\mathbf{x}_{stego}$
\end{algorithmic}
\end{algorithm}

By freezing the original weights of the conditional diffusion model and only updating the parameters of the visual path, agentic AI effectively preserves the original generative capabilities while embedding the specific characteristics of the reference image. The detailed training and inference processes are shown in \textbf{Algorithm \ref{alg:ipadapter_edict}}.

Combining the EDICT sampling and the decoupled cross-attention mechanism together, the sampling of stego image in the transmitter can be rewritten as a conditional process as:
\begin{equation}
\label{eq:edict_rev_ddimlike}
z_{t-1}
=\alpha_t\big(p\,\mathbf{z}_t+(1-p)\,\mathbf{u}_t\big)+\beta_t\,\tilde\epsilon_{t,c},
\end{equation}
\begin{equation}
\begin{aligned}
\label{eq:edict_rev_fused_eps}
\tilde\epsilon_{t,c}
& =p\,\epsilon_\theta(\mathbf{u}_t,t,K_{\text{pub}},\mathbf{x}_{ref}) \\ 
&\quad +(1-p)\,\epsilon_\theta(\mathbf{z}_t^{\mathrm{inter}},t,K_{\text{pub}},\mathbf{x}_{ref}),
\end{aligned}
\end{equation}
\begin{equation}
\label{eq:edict_rev_inter}
\mathbf{z}_t^{\mathrm{inter}}=\alpha_t \mathbf{z}_t+\beta_t\,\epsilon_\theta(\mathbf{u}_t,t,K_{\text{pub}},\mathbf{x}_{ref}),
\end{equation}
where $\tilde\epsilon_{t,c}$ is the noise prediction in $t$ step.

\section{Simulation Results and Analysis}

\subsection{Simulation Parameters}
\subsubsection{Datasets}
We use UniStega \cite{yang2024diffstega} datasets to verify the performance of AgentSemSteCom. UniStega is composed of three subsets designed to represent different semantic steganography scenarios, with a total of 100 images and relative prompts. The images are collected from multiple public datasets, including COCO \cite{lin2014microsoft}, AFHQ\cite{choi2020stargan}, and CelebA-HQ \cite{karras2017progressive}, as well as other online sources. All images are processed through center cropping and uniformly resized as 512$\times$512. The prompts are generated by BLIP \cite{li2022blip} and Llama2 \cite{touvron2023llama}.

In our experiment, we consider images of three different classes. We use the UniStega-Content to simulate the common steganography scene, which focuses on the natural images of general objects and backgrounds. UniStega-Style is used to simulate the stylized scene, which abstracts the real objects into artistic creations on specific themes. For facial scenes, which require high-precision recovery of human identities, we use facial-image subset from UniStega-Similar. We also make minor adjustments to some inaccurate public semantic keys.

\begin{table*}
\centering
\caption{Performance of AgentSemSteCom under Different classes of images and SNRs, where the MSE is scaled by $10^{-2}$.}
\label{tab:three_scenarios}
\setlength{\tabcolsep}{5pt}
\renewcommand{\arraystretch}{1.0}
\begin{tabular}{c cccc cccc cccc}
\toprule
\multirow{2}{*}{\textbf{SNR (dB)}} 
& \multicolumn{4}{c}{\textbf{Common steganography images}} 
& \multicolumn{4}{c}{\textbf{Facial images}} 
& \multicolumn{4}{c}{\textbf{Style images}} \\
\cmidrule(lr){2-5} \cmidrule(lr){6-9} \cmidrule(lr){10-13}
& PSNR & SSIM & MSE & LPIPS
& PSNR & SSIM & MSE & LPIPS
& PSNR & SSIM & MSE & LPIPS \\
\midrule
5  & 19.71~\text{dB} & 0.4778 & 1.207 & 0.5022
   & 22.74~\text{dB} & 0.6989 & 0.566 & 0.4042
   & 22.19~\text{dB} & 0.6233 & 0.721 & 0.3850 \\
10 & 21.59~\text{dB} & 0.5643 & 0.799 & 0.4030
   & 25.36~\text{dB} & 0.7938 & 0.303 & 0.2385
   & 24.54~\text{dB} & 0.7069 & 0.424 & 0.2832 \\
15 & 22.06~\text{dB} & 0.5880 & 0.713 & 0.3829
   & 26.16~\text{dB} & 0.8154 & 0.254 & 0.2127
   & 25.04~\text{dB} & 0.7248 & 0.378 & 0.2798 \\
20 & 22.15~\text{dB} & 0.5933 & 0.697 & 0.3821
   & 26.29~\text{dB} & 0.8186 & 0.247 & 0.2127
   & 25.16~\text{dB} & 0.7288 & 0.368 & 0.2828 \\
\bottomrule
\end{tabular}
\end{table*}

\begin{figure}
\centering
\includegraphics[width=0.9\linewidth]{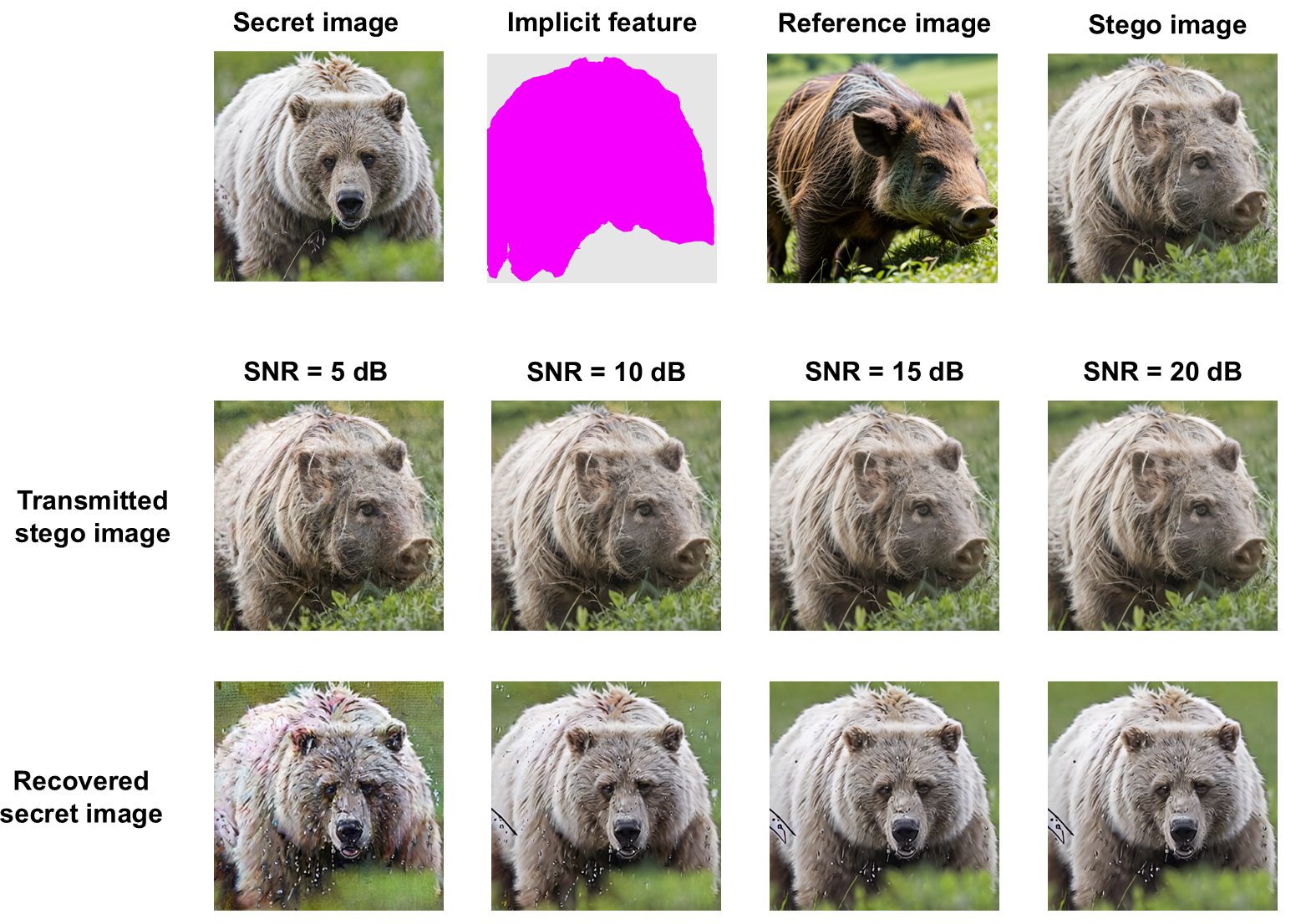}
\caption{Visualization results of recovery
images over different SNRs, where the public key is “a wild boar walking through a lush green field".}
\label{different_snrs}
\end{figure}
\subsubsection{Parameter Settings}
In the AgentSemSteCom, all the modules are pre-trained, with the detailed parameters as follows.
\begin{itemize}
\item 
\textbf{The Implicit Feature Extraction Module:} The OneFormer model \cite{jain2023oneformer} is employed to output semantic segmentation map as the implicit feature in the common scenario. The OpenPose model \cite{simon2017hand} is employed to generate skeleton in the facial scenario. The style scenario is only controlled by the public semantic key without any implicit feature.
\item 
\textbf{The Reference Image Generation Module:} We employ PicX-Real\footnote{\url{https://huggingface.co/GraydientPlatformAPI/picx-real}}  as an independent diffusion-based generator. Unlike the diffusion model used in the steganographic hiding and recovery processes, PicX-Real is specifically adopted to synthesize reference images with stable visual appearance and high realism. As the reference image is jointly controlled by the extracted implicit feature and the digital token, ControlNet is employed to inject the implicit feature into the diffusion process to explicitly impose structural constraints.
\item 
\textbf{The Coverless Steganography Module:} We use the Stable Diffusion v1.5\footnote{\url{https://huggingface.co/stable-diffusion-v1-5/stable-diffusion-v1-5}} in the forward and reverse diffusion process, with the 50 EDICT steps and 0.05 noise perturbation strength $\eta$. The mixing coefficient of EDICT is set to 0.93. Also, we introduce a parameter, edit strength, to define the proportion of diffusion steps applied during the reverse diffusion stages. In the practical simulation, we set it to 0.5 to balance the semantic steganographic effect and numerical drift, which is caused by the accumulation of floating-point rounding errors in diffusion steps. The IPAdapter-plus is employed for guidance injection of reference image with its weight factor set to 1.
\item 
\textbf{The Semantic Transmission Module:} AgentSemSteCom is verified under the AWGN channel ranging from 5 dB to 20 dB SNR in both analog and digital SemCom framework, where the coefficient of physical channel $h$ is set to 1. For the analog SemCom, we use  SwinJSCC\cite{yang2024swinjscc} as backbone trained on the CelebA-HQ dataset under an additive white Gaussian noise (AWGN) channel. The channel bandwidth parameter is set to C=48, and the trained SNR is fixed at 10 dB. For the digital SemCom framework, we utilize MDJCM \cite{zhang2024analog} trained on DIV2K dataset \cite{agustsson2017ntire} with modulation order of 64 and SNR at 10~dB.
\item 
\textbf{The Task-oriented Enhancement Module:} We employ the CodeFormer \cite{zhou2022towards} to enhance the texture and detail of facial image. This module leverages a learned discrete codebook prior to cast face restoration as a code prediction task, which relies on a transformer-based prediction network. The model is trained on FFHQ \cite{karras2019style} dataset with 512*512 input. In our experiment, the fidelity weight is set to 0.5 to balance the perceptual quality and identity fidelity.
\end{itemize}
\subsection{Evaluation Metrics and Comparison Schemes}
\subsubsection{Evaluation Metrics}
To comprehensively evaluate the performance of AgentSemSteCom, we employ PSNR and MSE to quantify pixel-level reconstruction accuracy, while SSIM evaluates structural consistency between the reconstructed image and the reference. LPIPS \cite{zhang2018unreasonable} is further employed to measure perceptual similarity in deep feature space. 
\subsubsection{Comparison Scheme}
To validate the effectiveness and superiority of AgentSemSteCom, we compare it with SemSteDiff \cite{gao2025semstediff}. It is noted that, SemSteDiff requires updating the private semantic key for each transmitted image.
\subsection{Simulation Results}
\subsubsection {Performance of the Proposed AgentSemSteCom Scheme}
\paragraph{Generalization performance under different image classes}
To measure the effectiveness of AgentSemSteCom, we evaluate it in images of three classes mentioned before, under $\text{SNR}=5, 10, 15, 20~\text{dB}$ respectively. Table \ref{tab:three_scenarios} shows the quantitative results of four metrics. In general, AgentSemSteCom maintains stable pixel-level reconstruction accuracy across different task requirements and SNRs, for almost all PSNR exceeds 20 dB and MSE remains on the order of $10^{-2}$. At $\text{SNR}=5~\text{dB}$,  the facial class demonstrates remarkable robustness, achieving a PSNR of $22.73~\text{dB}$ and an SSIM of $0.6989$. Also, the style class manages to keep the LPIPS at a low $0.3850$ and MSE at $0.721 \times 10^{-2}$, showcasing the effectiveness of agentic reasoning in preserving hidden intent under severe channel distortion. As the channel quality improves toward $\text{SNR}=20~\text{dB}$, the metrics across all classes converge to high-fidelity reconstruction, with the facial class reaching a peak PSNR of $26.29~\text{dB}$ and a significantly refined LPIPS of $0.2127$. The positive correlation between SNR growth and metric improvement is deeply related to noise-sensitive EDICT-based diffusion process. During the semantic feature extraction and channel transmission, certain high-frequency components of the secret image may be ignored or distorted, which influence the deterministic reverse trajectory. Figure \ref{different_snrs} shows visualized comparison results of recovery images across different SNRs.

\begin{figure*}
\centering
\includegraphics[width=\linewidth]{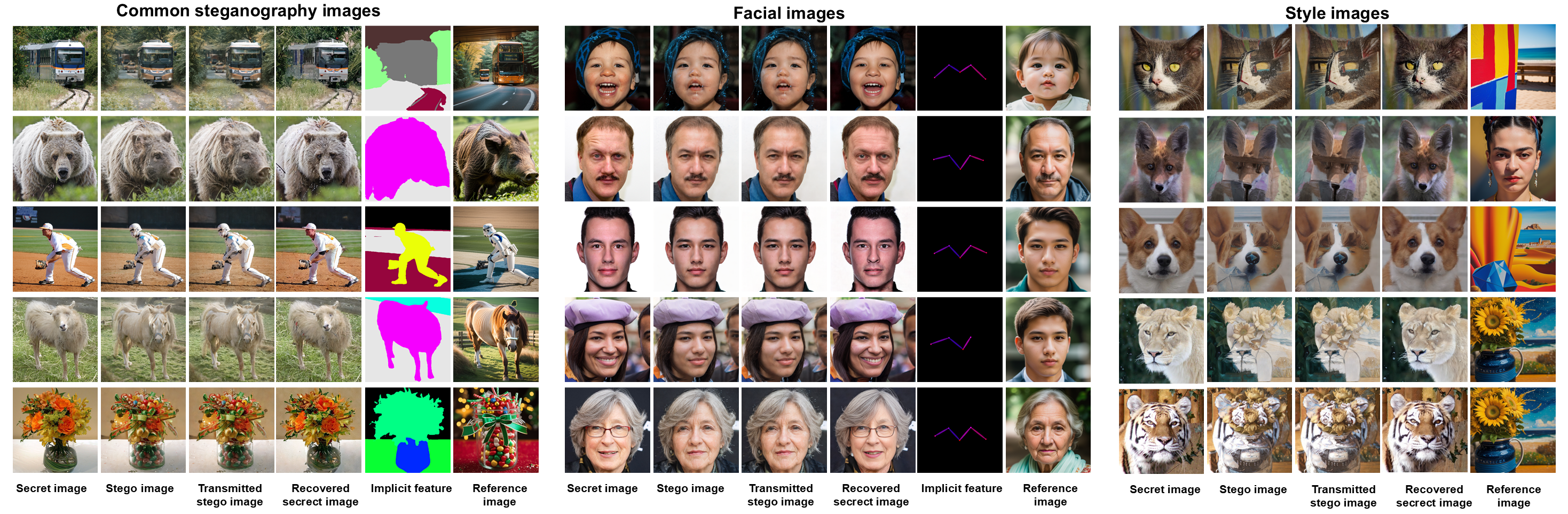}
\caption{Visualization results of AgentSemSteCom under different classes of images, which include common steganography images, facial images and style images. The images are collected at $\text{SNR = 10~dB}$.}
\label{scenario}
\end{figure*}

\begin{figure*}[]
    \centering
    \begin{subfigure}[b]{0.24\textwidth}
        \includegraphics[width=\linewidth]{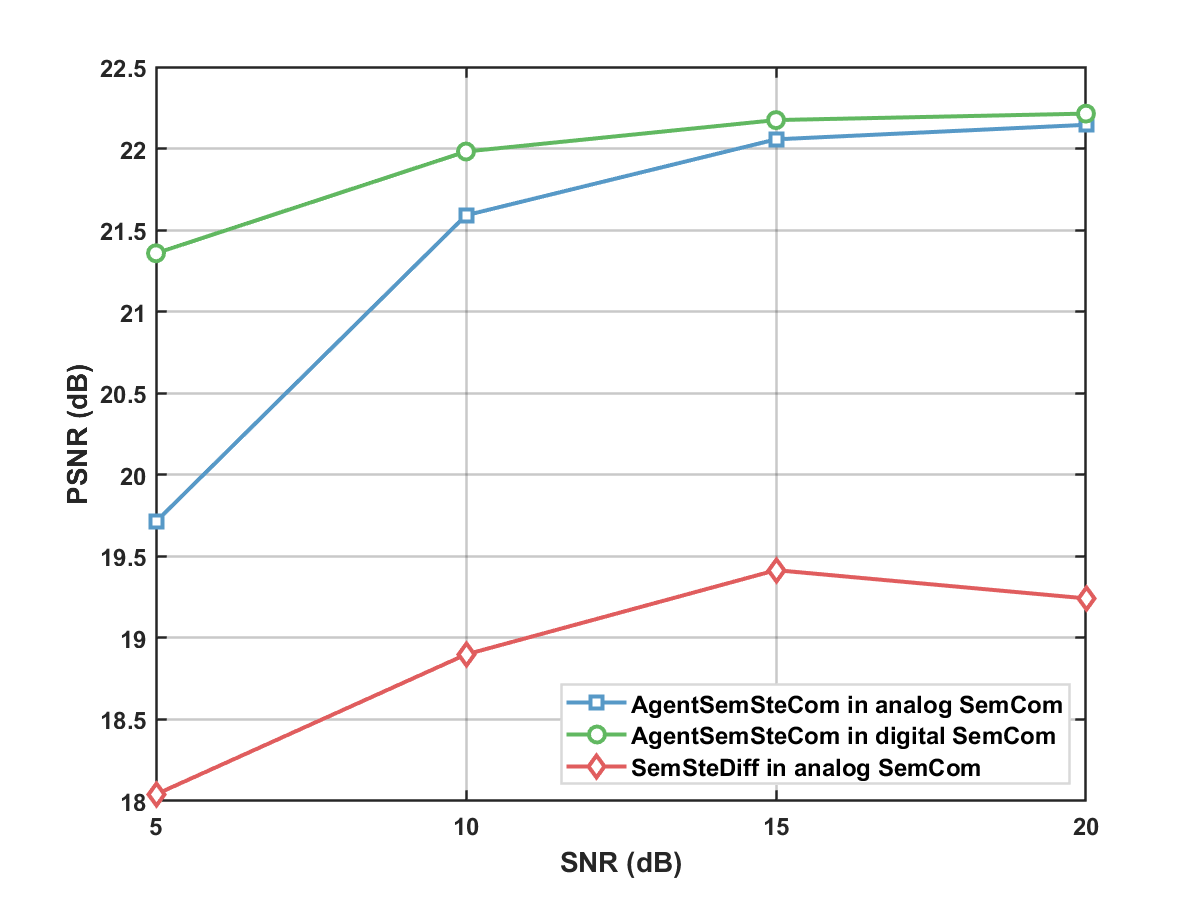}
        \caption{Comparison on PSNR}
        \label{comparePSNR}
    \end{subfigure}
    \hfill
    \begin{subfigure}[b]{0.24\textwidth}
        \includegraphics[width=\linewidth]{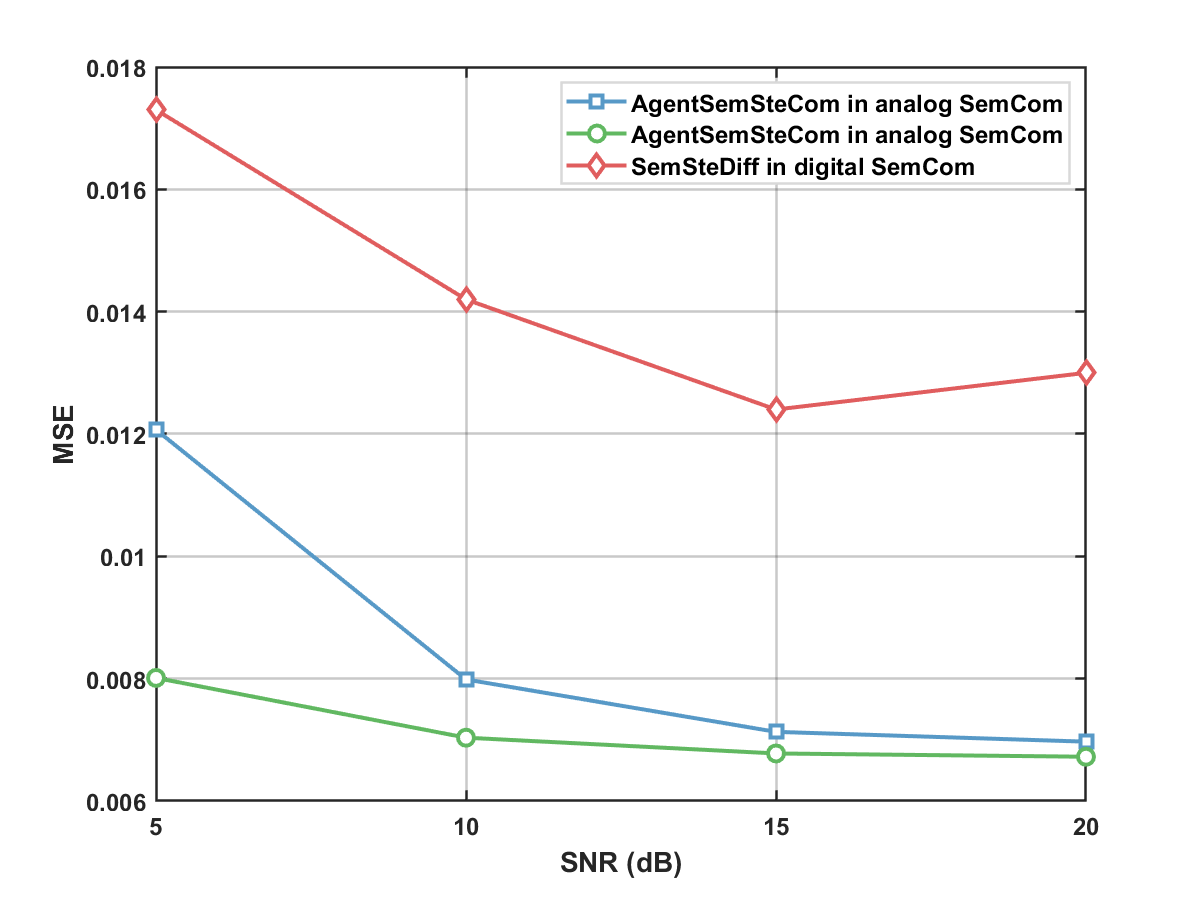}
        \caption{Comparison on MSE}
        \label{compareMSE}
    \end{subfigure}
    \hfill
    \begin{subfigure}[b]{0.24\textwidth}
        \includegraphics[width=\linewidth]{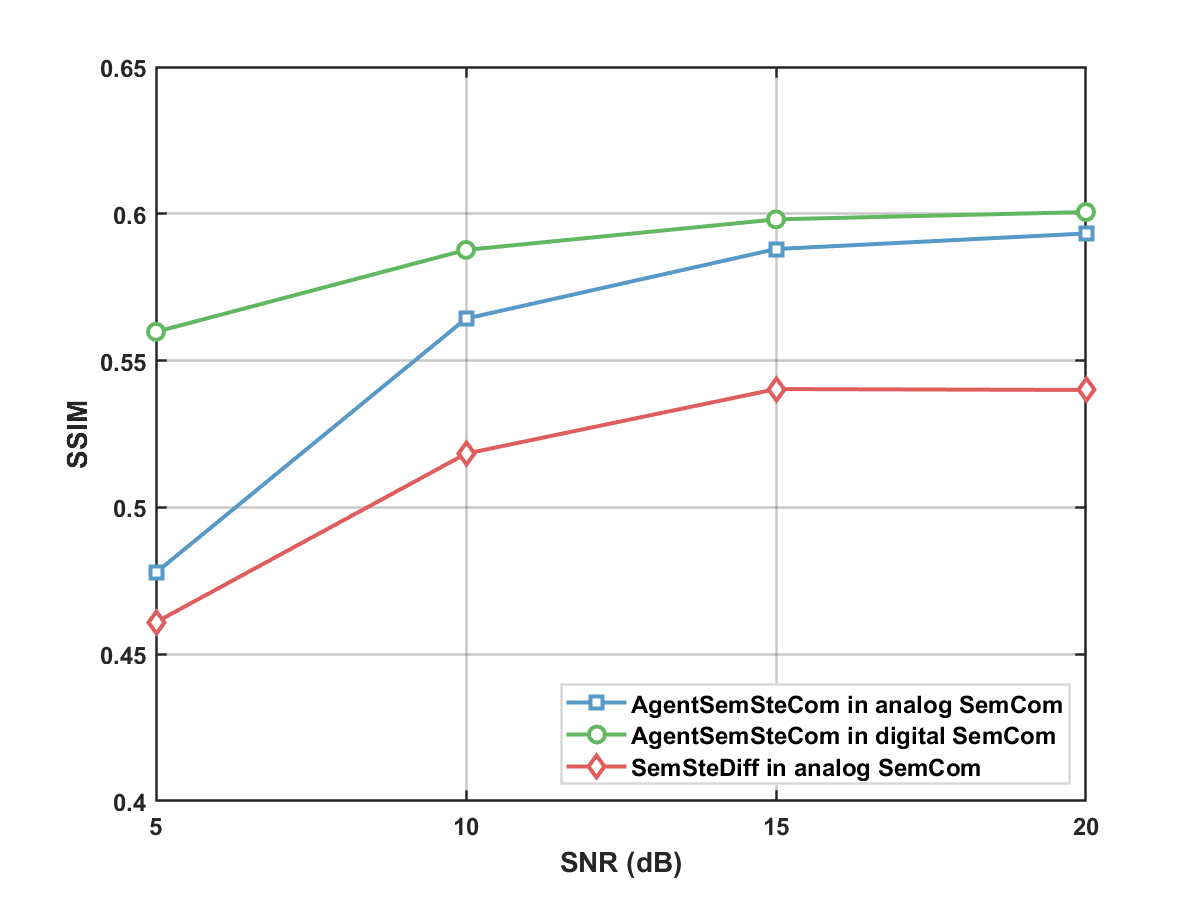}
        \caption{Comparison on SSIM}
        \label{compareSSIM}
    \end{subfigure}
    \hfill
    \begin{subfigure}[b]{0.24\textwidth}
        \includegraphics[width=\linewidth]{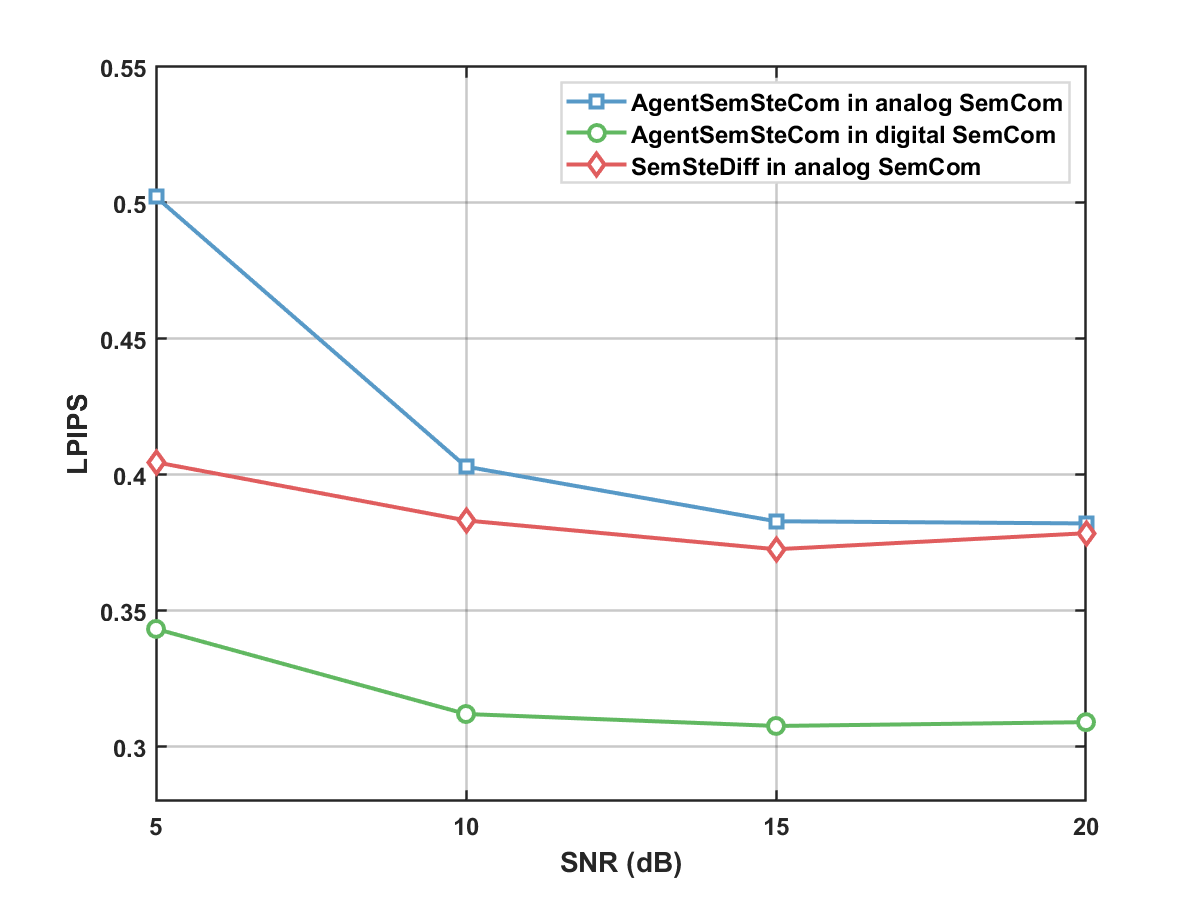}
        \caption{Comparison on LPIPS}
        \label{compareLPIPS}
    \end{subfigure}
    \caption{Comparison between AgentSemSteCom and SemSteDiff, which is simulated towards common steganography images.}
    \label{fig:compare}
\vspace{-10pt}
\end{figure*}
Moreover, the performance difference among image classes is caused by the different implicit features. As shown in Table \ref{tab:three_scenarios}, the facial class consistently achieves the best reconstruction quality across all SNR levels, for the OpenPose-based skeletal landmarks provides geometric structure among facial components. Such constraints enable the EDICT-based diffusion process to accurately preserve identity-related spatial configurations, resulting in higher PSNR and SSIM and lower MSE. In comparison, the common steganography class shows relatively lower reconstruction result, which is limited by semantic segment image that captures global semantic region rather than detailed local structures. The style class shows the performance without reference image guidance, the PSNR is comparable to the counterpart of facial class. This indicates that the LLM-generated public semantic key is sufficient for AgentSemSteCom in specific situation. To better understand the effect of different implicit feature, Figure \ref{scenario} shows the visualization of steganography across three classes at $\text{SNR}=10~\text{dB}$.

\paragraph{Performance comparison with SemSteDiff}
\begin{figure}[t]
\centering
\includegraphics[width=0.9\linewidth]{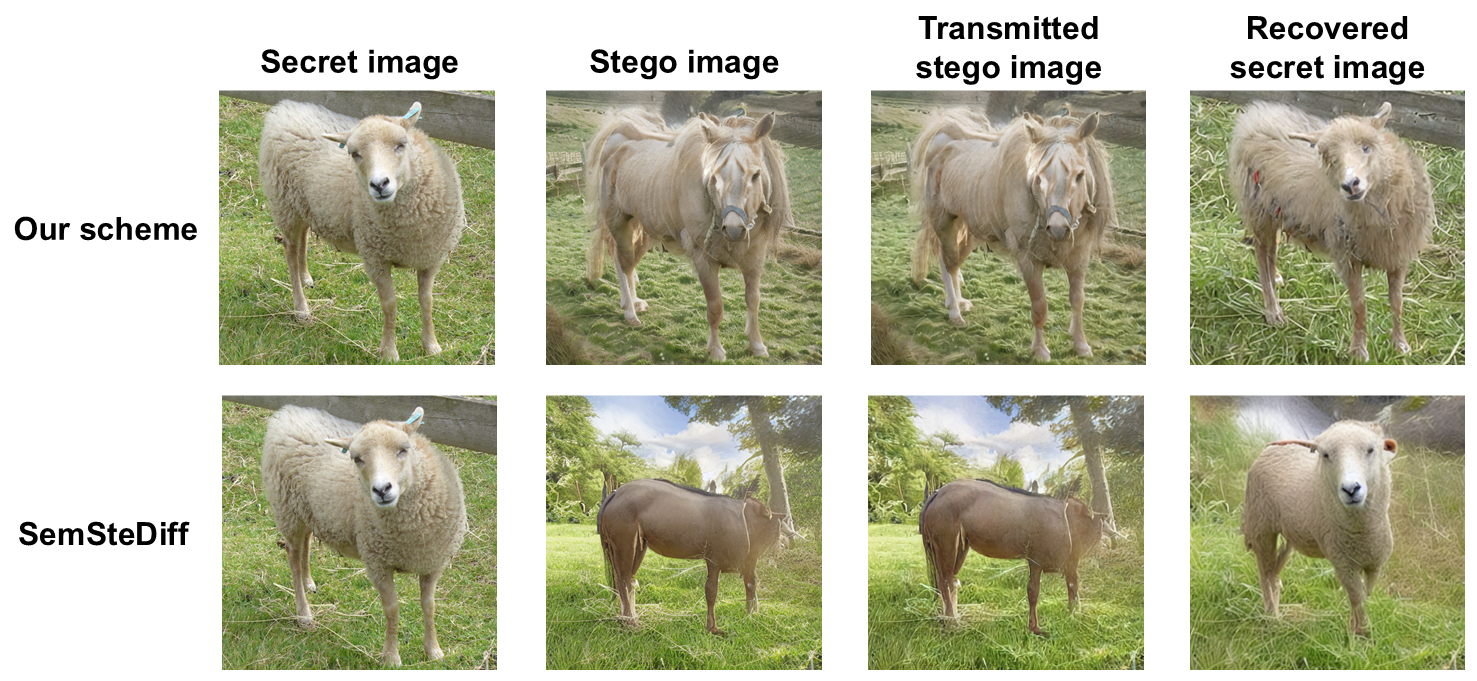}
\caption{Visualization results of AgentSemSteCom and SemSteDiff, where the private key for SemSteDiff is ``a sheep standing in a field next to a wooden fence", the public key for both schemes is ``a horse standing in a field next to a wooden fence". The comparison images are collected at $\text{SNR = 10~dB}$.}
\label{comparision}
\end{figure}
Figure \ref{fig:compare} illustrates the performance of AgentSemSteCom across digital and analog SemCom frameworks, and the comparison between our scheme and SemSteDiff in analog SemCom towards image of common steganography class. In terms of pixel-level reconstruction accuracy, the PSNR and MSE curves demonstrate that AgentSemSteCom significantly outperforms the SemSteDiff baseline. Specifically, at $\text{SNR}=5~\text{dB}$, AgentSemSteCom achieves a PSNR of $19.71~\text{dB}$ and a lower MSE of $1.207 \times 10^{-2}$, while SemSteDiff maintains $18.04~\text{dB}$ and $1.73 \times 10^{-2}$. This effect is primarily due to the EDICT-based exact inversion process, which effectively mitigates cumulative sampling errors and enables the system to mathematically restore the original secret image. In contrast, the previous SemSteDiff can only generate semantic-similar images guided by public and private semantic keys, lacking the precision for exact reconstruction, which is clearly displayed in Figure \ref{comparision}. For the perception, 
AgentSemSteCom achieves a higher structural similarity, with an SSIM of 0.5643 at 10 dB, compared to SemSteDiff, which has an SSIM of 0.5183. The improved 
performance is attributed to the OneFormer-based segmentation maps, which guide the recovered image aligning with the original layout. However, such strong structural constraints can overly restrict the generative flexibility of the diffusion process, which may introduce minor local inconsistencies in visual appearance. As indicated by the LPIPS curves, SemSteDiff’s pure semantic guidance allows for more harmonious image synthesis without rigid structural boundaries, resulting in a lower perceptual distance of $0.4046$ compared to AgentSemSteCom of $0.5022$ at $\text{SNR}=5~\text{dB}$.

\begin{figure*}[]
    \centering
    \begin{subfigure}[b]{0.24\textwidth}
        \includegraphics[width=\linewidth]{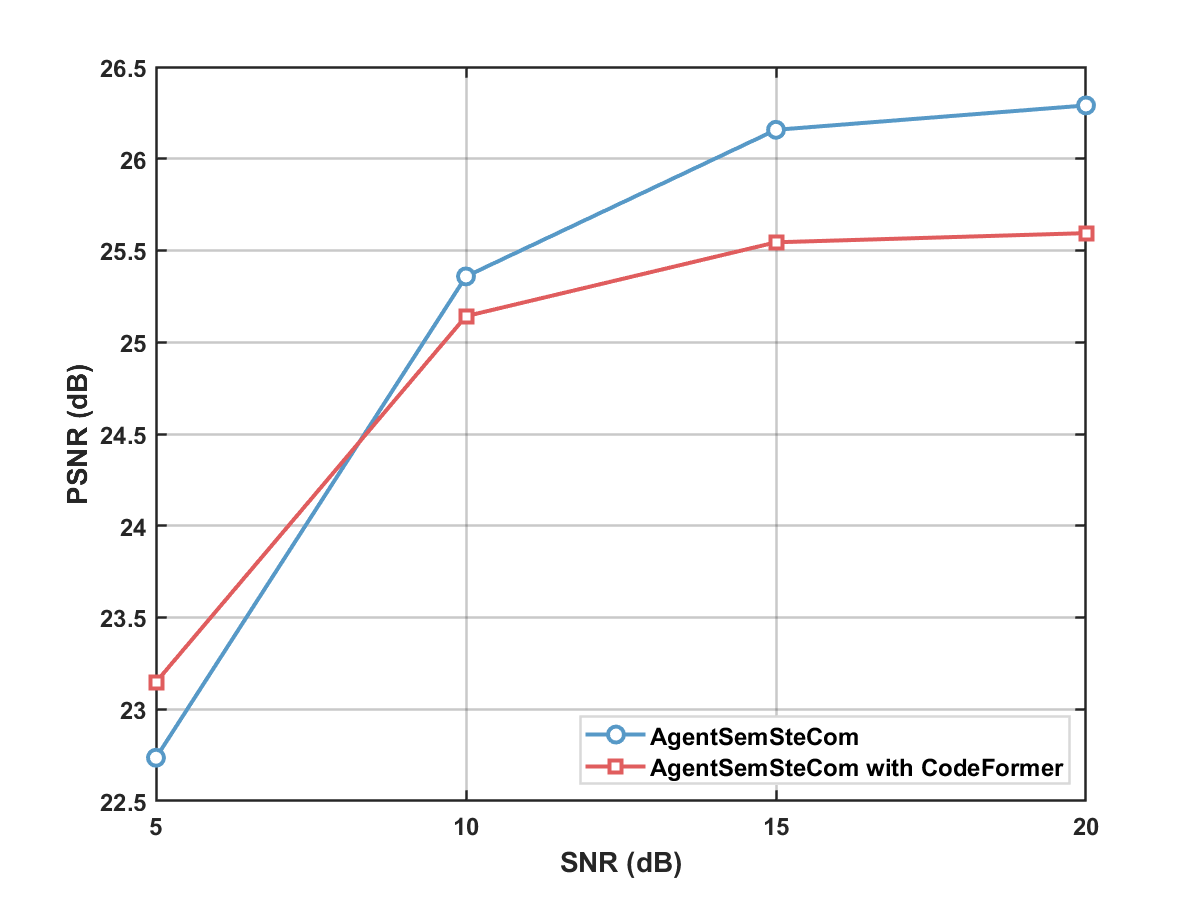}
        \caption{PSNR of CodeFormer}
        \label{enhanced_PSN}
    \end{subfigure}
    \hfill
    \begin{subfigure}[b]{0.24\textwidth}
        \includegraphics[width=\linewidth]{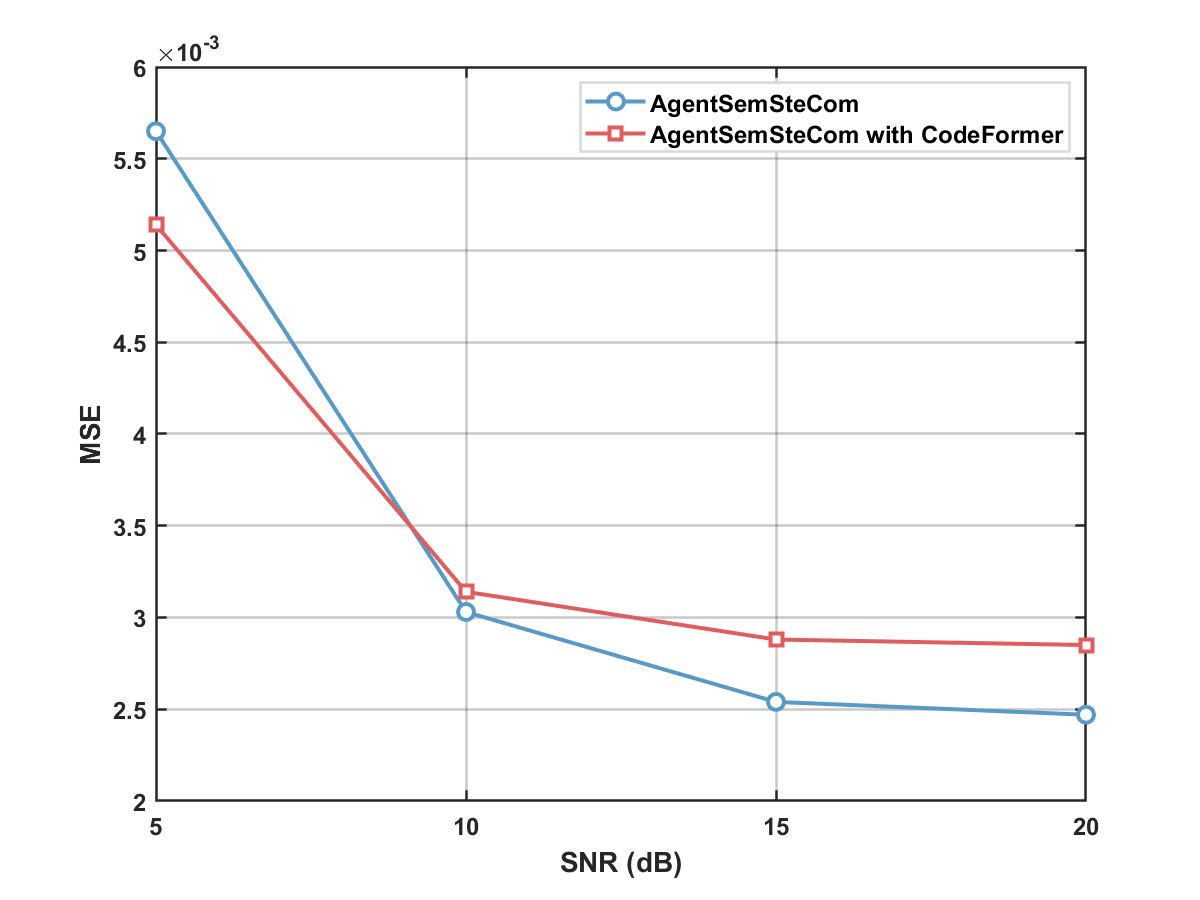}
        \caption{MSE of CodeFormer}
        \label{enhanced_MSE}
    \end{subfigure}
    \hfill
    \begin{subfigure}[b]{0.24\textwidth}
        \includegraphics[width=\linewidth]{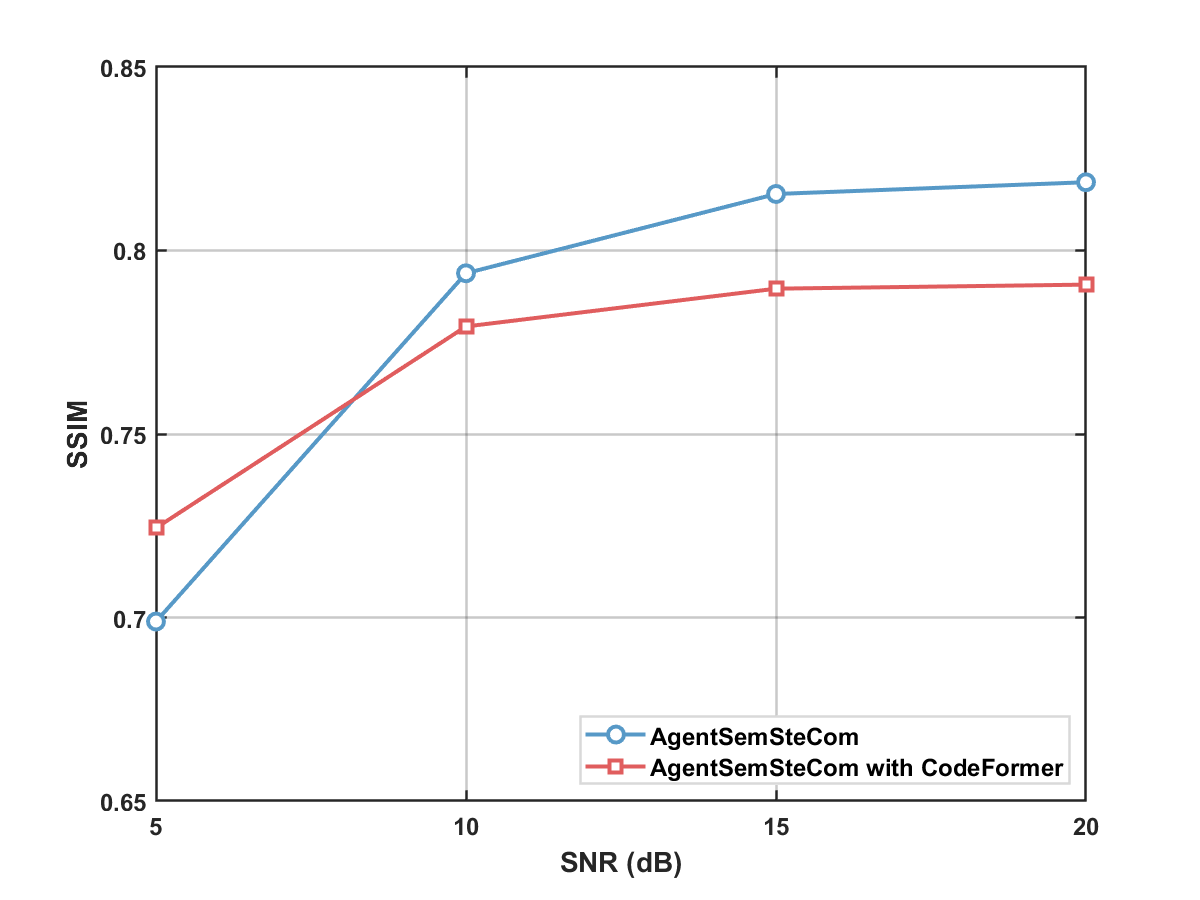}
        \caption{SSIM of CodeFormer}
        \label{enhanced_SSIM}
    \end{subfigure}
    \hfill
    \begin{subfigure}[b]{0.24\textwidth}
        \includegraphics[width=\linewidth]{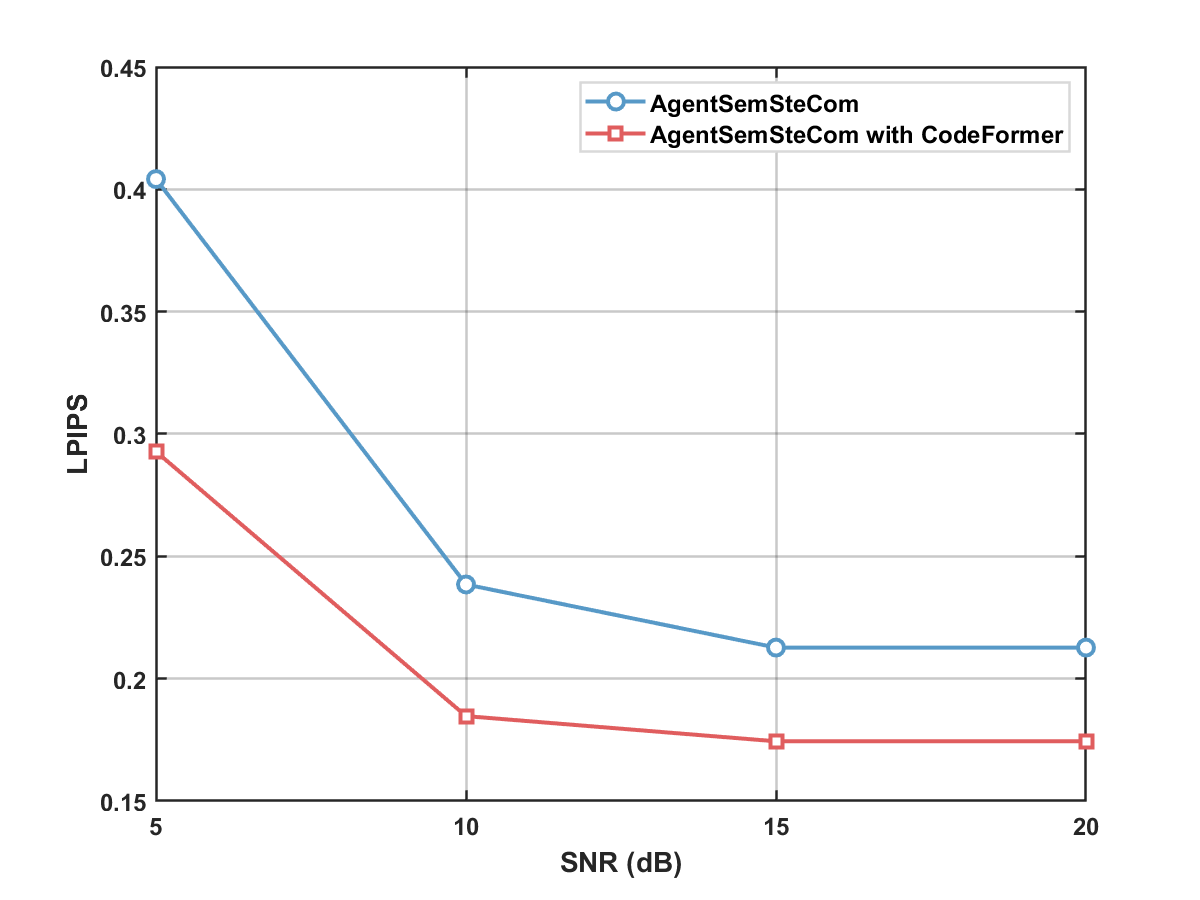}
        \caption{LPIPS of CodeFormer}
        \label{enhanced_LPIPS}
    \end{subfigure}
    \caption{Performance improvement of AgentSemSteCom with the task-oriented enhancement module, which is simulated towards facial images.}
    \label{figure:enhance}
    \vspace{1mm} 
\end{figure*}
\begin{figure}[]
\centering
\includegraphics[width=0.8\linewidth]{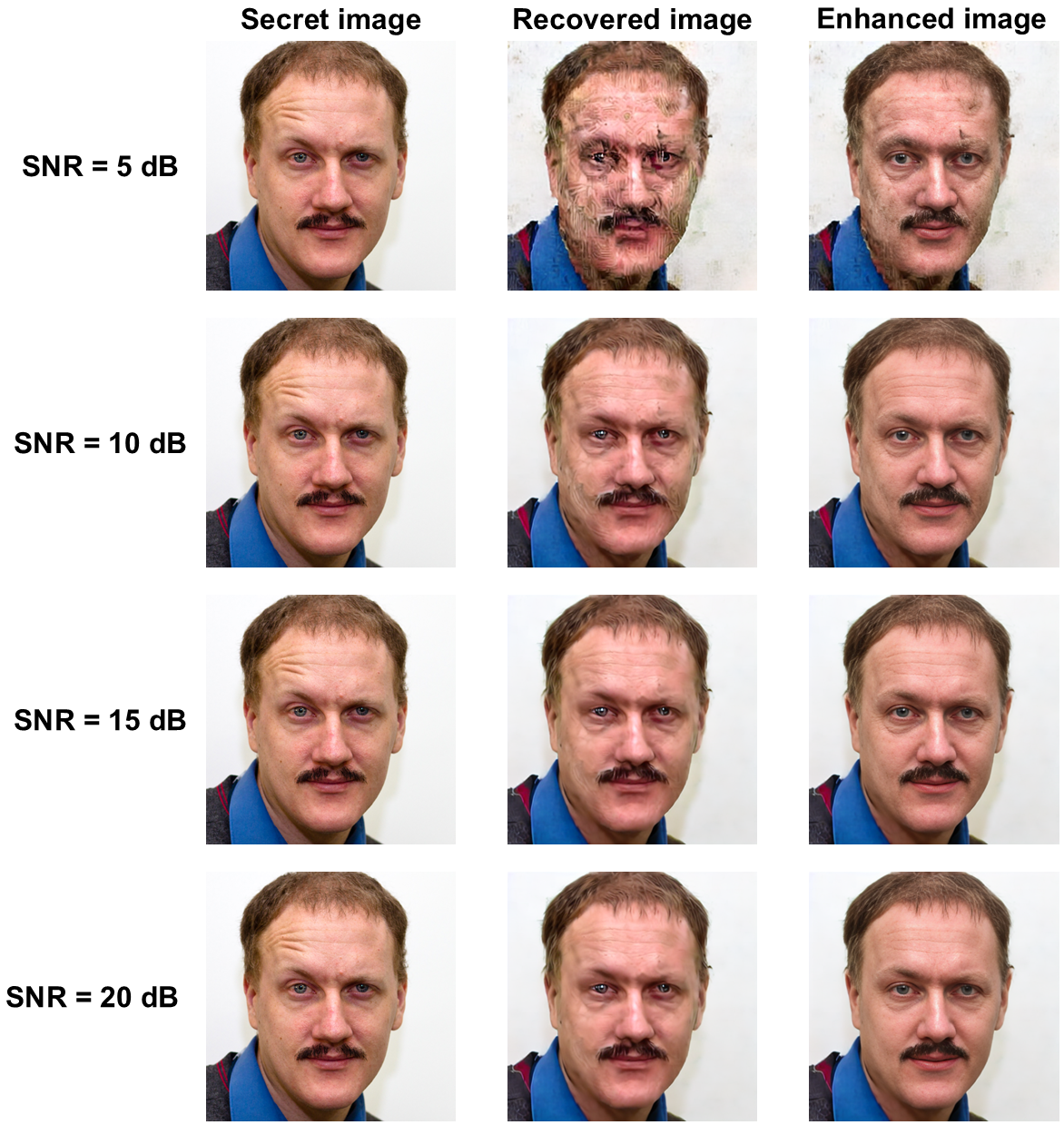}
\caption{Visualization results of AgentSemSteCom with the task-oriented enhancement module under different SNRs.}
\label{enhance}
\end{figure}

\paragraph{Performance improvement of optional task-oriented enhancement module}
Figures \ref{figure:enhance} and \ref{enhance} present the performance improvement of AgentSemSteCom in the facial class following the integration of the CodeFormer restoration module. This optional task-oriented enhancement module demonstrates the intelligent adaptability of our agentic framework, where the receiver autonomously invokes specialized tools to complement semantic reconstruction in the low-SNR region. At $\text{SNR}=5~\text{dB}$, the improvement is comprehensive across all evaluation metrics: the PSNR and SSIM increase from $22.74~\text{dB}$ and $0.6989$ to $23.15~\text{dB}$ and $0.7244$, while the MSE and LPIPS decrease from $0.566 \times 10^{-2}$ and $0.4042$ to $0.514 \times 10^{-2}$ and $0.2928$, respectively. As the SNR increases, the enhancement module primarily contributes to perceptual improvement, while its impact on pixel-level fidelity becomes limited. For instance, at $\text{SNR}=20$ dB, the enhanced PSNR reaches $25.59$ dB, which is lower than the $26.29$ dB achieved by baseline. The LPIPS value consistently performs better with the enhancement module. This divergence depends on the generative nature of CodeFormer, which leverages a discrete codebook prior to emphasize perceptual realism rather than strict pixel-wise consistency with the original secret image.




\subsubsection {Security Evaluation}
\paragraph{The visualization of stego images with different digital tokens}

\begin{figure}
\centering
\includegraphics[width=0.8\linewidth]{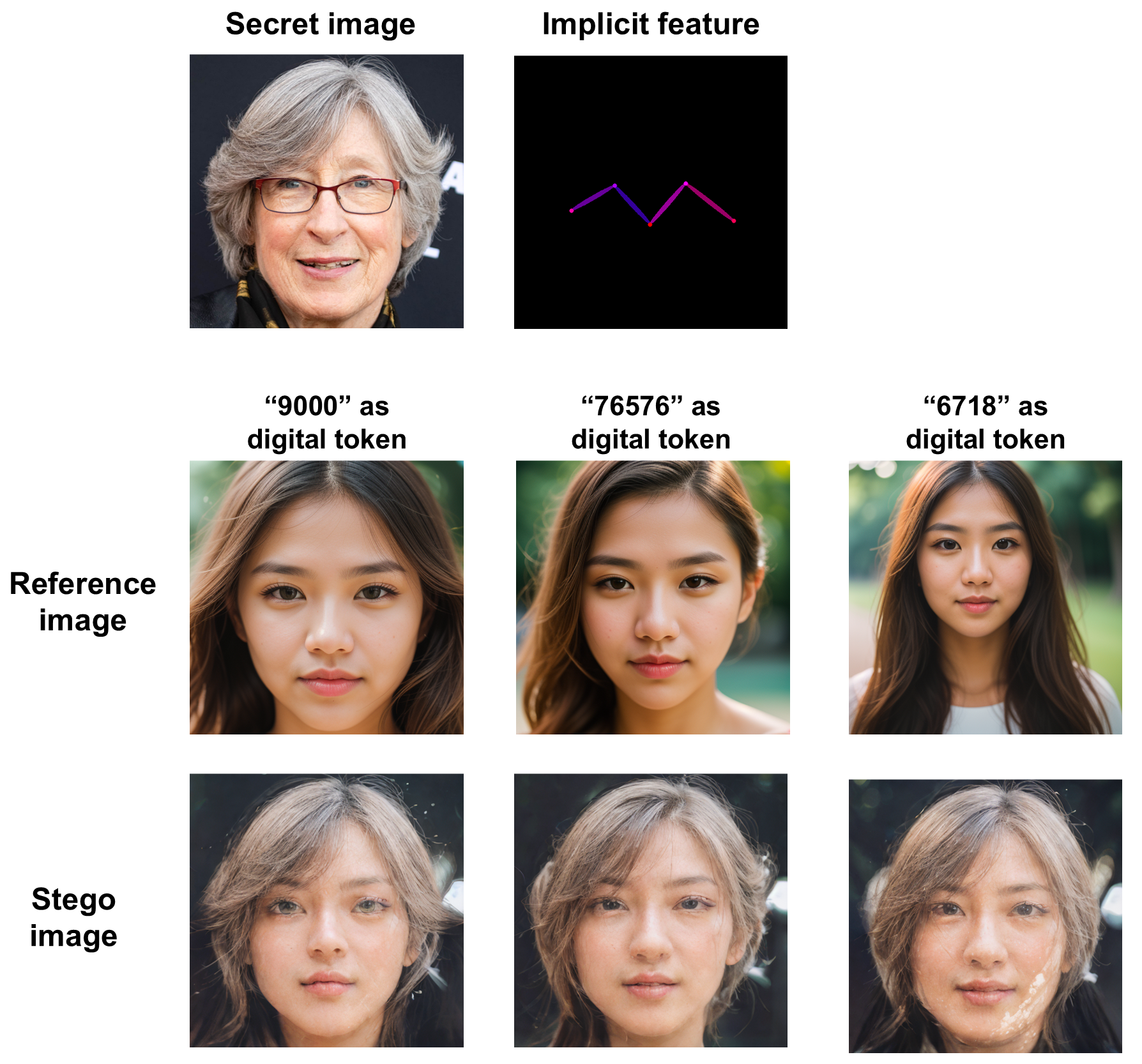}
\caption{Visualization of stego image generation with different digital tokens.}
\label{tokens}
\end{figure}

The security of the proposed AgentSemSteCom framework first enables by the digital token. A single secret image can be mapped to multiple visually distinct stego images by varying the token, without introducing any other information. Specifically, Figure \ref{tokens} shows the result of different digital tokens including 9000, 76576, and 6718 at $\text{SNR}=10$ dB, with edit strength of $0.5$, under PicX-Real model. The stego images exhibit noticeable differences in local facial attributes such as eye shape, nose geometry, and mouth contours, but the global structural configuration remains stable due to the landmark-based constraints. Moreover, agentic AI can further increase this diversity by autonomously selecting style-adaptive generative tools from a shared model library. This design allows users to generate different steganographic outputs from identical source content while the absence of explicit private semantic keys avoids suspicious key exchange behavior, which maintains a private keyless communication paradigm. 

\paragraph{Security evaluation under eavesdropping scenarios}
\begin{figure*}[]
    \centering
    \begin{subfigure}[b]{0.24\textwidth}
        \centering
        \includegraphics[width=\linewidth]{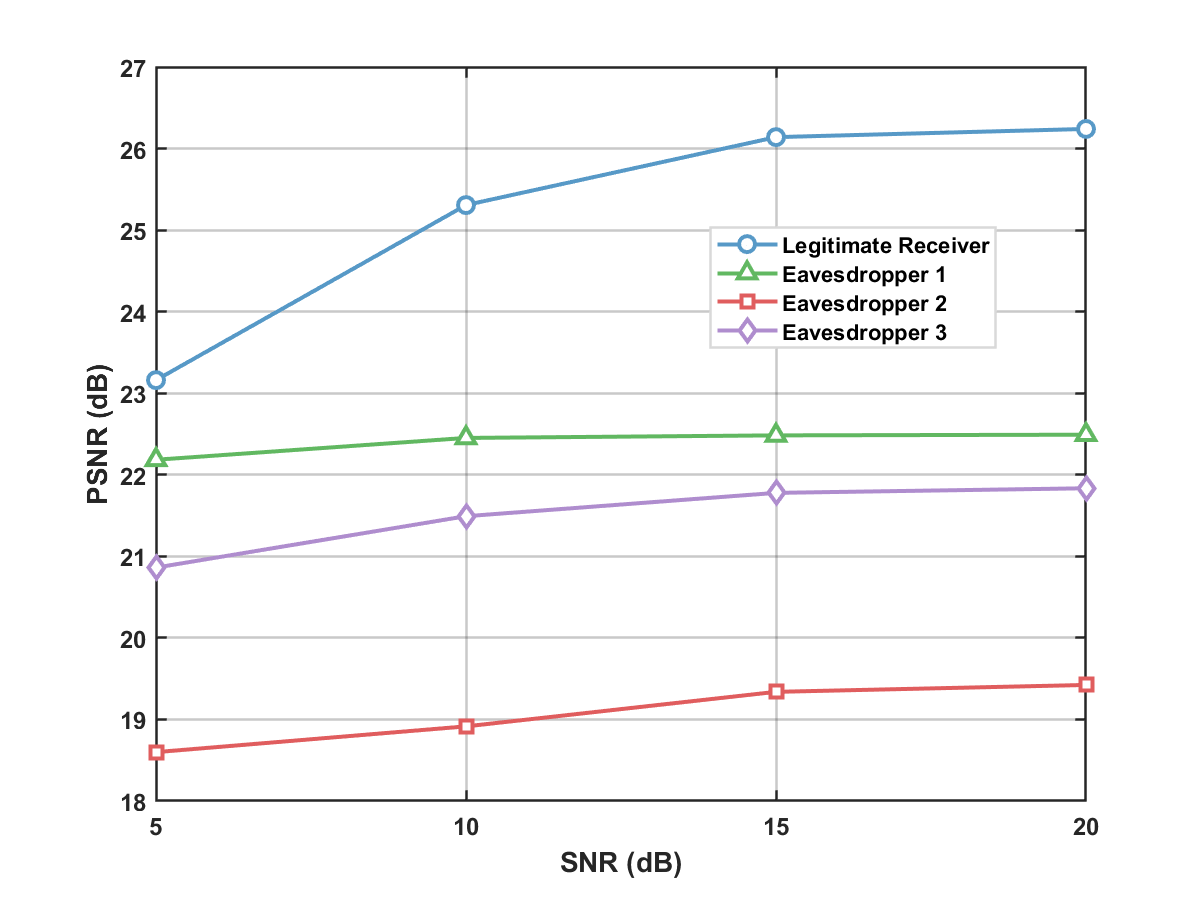}
        \caption{PSNR of eavesdroppers}
        \label{attackerPSNR}
    \end{subfigure}
    \hfill
    \begin{subfigure}[b]{0.24\textwidth}
        \centering
        \includegraphics[width=\linewidth]{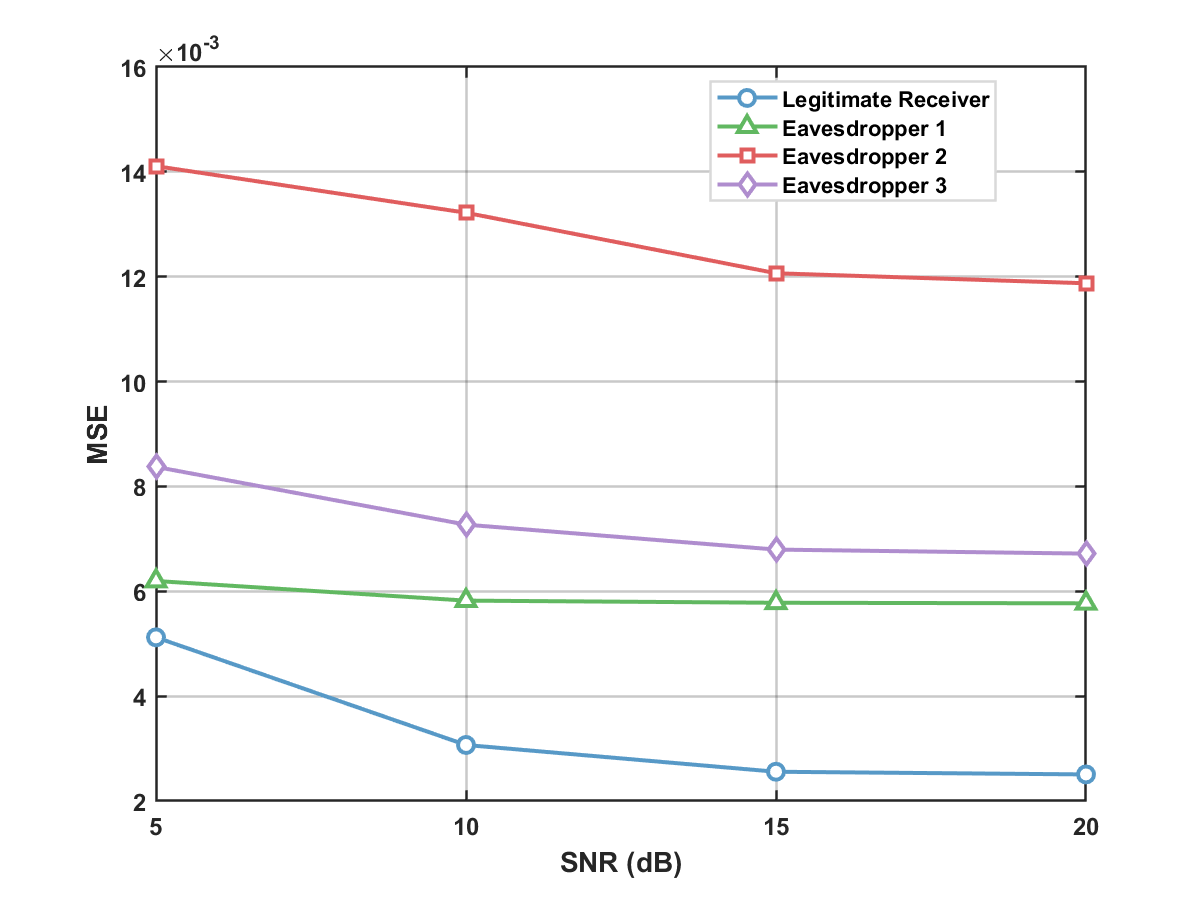}
        \caption{MSE of eavesdroppers}
        \label{attackerMSE}
    \end{subfigure}
    \hfill
    \begin{subfigure}[b]{0.24\textwidth}
        \centering
        \includegraphics[width=\linewidth]{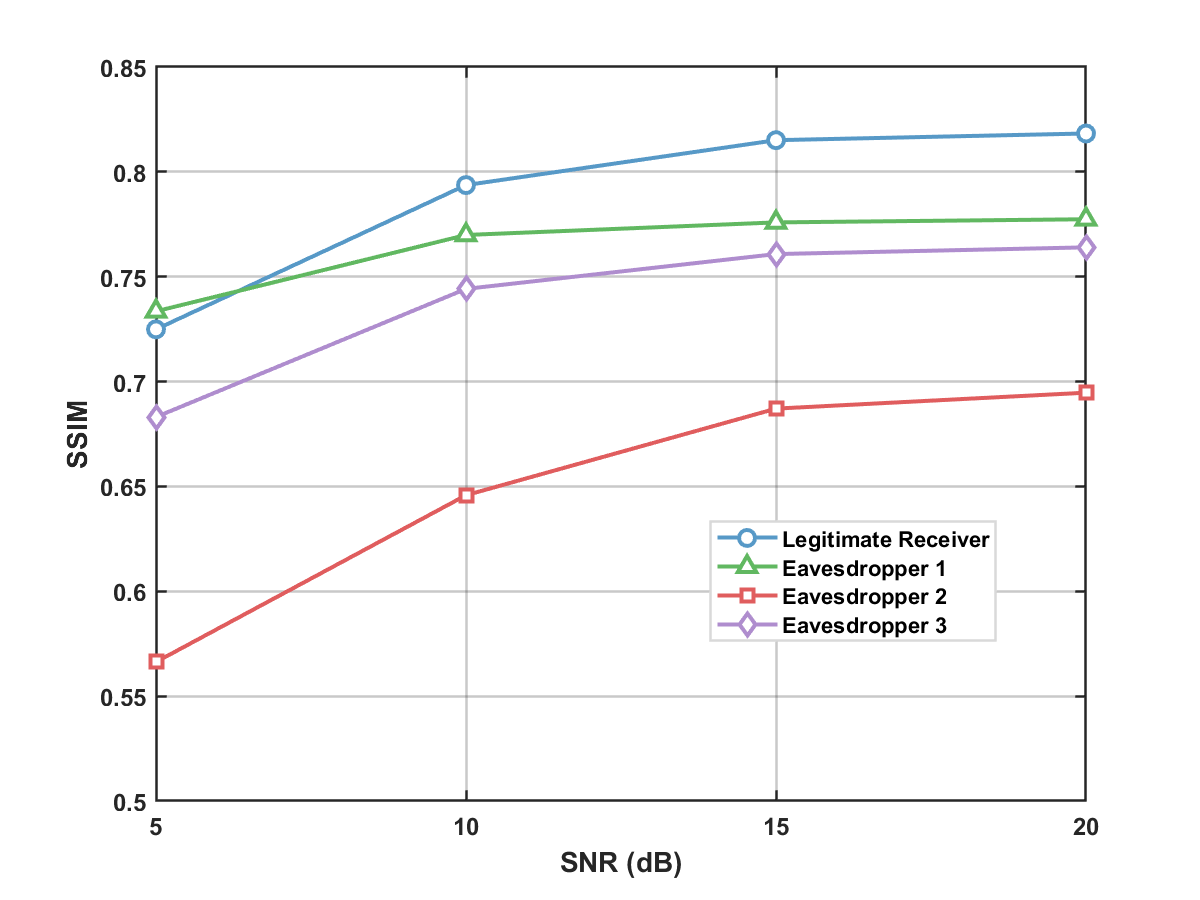}
        \caption{SSIM of eavesdroppers}
        \label{attackerSSIM}
    \end{subfigure}
    \hfill
    \begin{subfigure}[b]{0.24\textwidth}
        \centering
        \includegraphics[width=\linewidth]{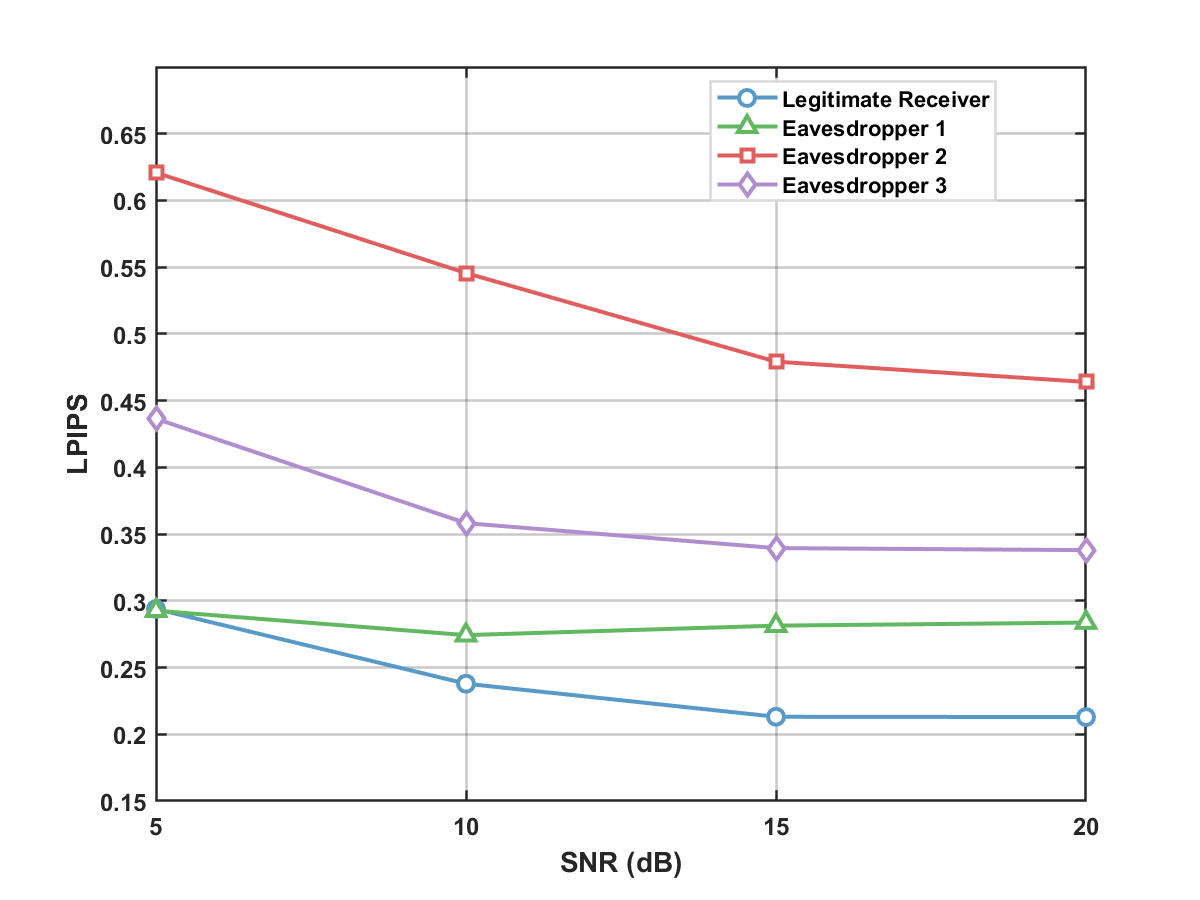}
        \caption{LPIPS of eavesdroppers}
        \label{attackerLPIPS}
    \end{subfigure}
    
    \caption{Comparison between legitimate receiver and eavesdroppers, which is simulated towards facial images.}
    \label{fig:attacker}
\end{figure*}
\begin{figure*}
\centering
\includegraphics[width=\linewidth]{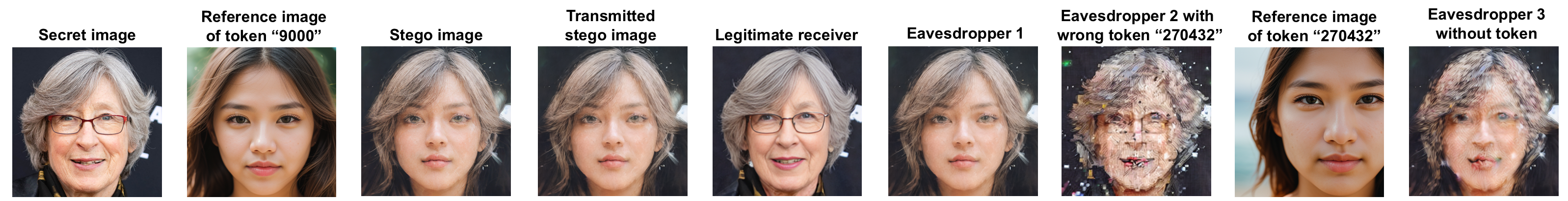}
\caption{Visualization results of legitimate receiver and eavesdroppers.}
\label{evasdropper}
\end{figure*}

To further assess the security performance of AgentSemSteCom, we model two categories of eavesdroppers, common eavesdroppers who intercept channel information to attempt traditional cracking, and agentic AI-based eavesdroppers capable of invoking local generative models to reconstruct the hidden content. We consider the worst case scenario for our evaluation, assuming the eavesdroppers have specific parameters of the legitimate semantic decoder, which is typically difficult to obtain in practical wireless environments. For agentic AI-based eavesdropper, we assume that both the public semantic key $K_{\text{pub}}$ and the implicit feature $\mathbf{x}_{feat}$ are leaked, and eavesdroppers have accurately inferred the implementation logic of our framework. Under these assumptions, three eavesdroppers with different capabilities are modeled as follows:
\begin{itemize}
\item \textbf{Eavesdropper 1:} The intelligent eavesdropper only possesses the semantic decoder without any key information. 
\item \textbf{Eavesdropper 2:} The agentic AI-based eavesdropper possesses the semantic decoder and the public semantic key and implicit feature, attempting to recover the secret image using an incorrect digital token and the available pre-trained LAMs.
\item \textbf{Eavesdropper 3:} The agentic AI-based eavesdropper possesses the semantic decoder and  the public semantic key and implicit feature, attempting to recover the secret image without any digital token by directly invoking pre-trained LAMs.
\end{itemize}

Figure \ref{fig:attacker} shows the quantitative curves of legitimate receiver and eavesdroppers towards images of facial class. Compared to the legitimate receiver, a significant performance gap is observed across all eavesdropping scenarios. Eavesdropper 2, attempting to recover the image with an incorrect token, suffers the most severe performance degradation, with a PSNR as low as $18.60~\text{dB}$ at $\text{SNR}=5~\text{dB}$ while the legitimate receiver achieves 23.16 dB. Eavesdropper 3, who attempts recovery without any token, performs slightly better at 20.86 dB than Eavesdropper 2. As SNR increases, the PSNR of legitimate receiver exceeds 26 dB, whereas agentic AI-based eavesdroppers remain around 20 dB, which strongly indicates that a wrong token causes the EDICT process to collapse into a chaotic state. The MSE, SSIM and LPIPS support the same conclusion. 
Eavesdropper 1 exhibits similar trends in PSNR and MSE as the other eavesdroppers. However, a slight gap between the legitimate receiver and Eavesdropper 1 is observable in SSIM and LPIPS, with Eavesdropper 1 unexpectedly achieving higher structural similarity and lower perceptual distance than the legitimate receiver. Specifically, its SSIM reaches $0.7335$, outperforming the legitimate receiver’s value of $0.7249$, while its LPIPS is reduced to $0.2926$, similar to $0.2939$ of receiver. This phenomenon validates the effectiveness of our steganographic scheme, for the eavesdropper 1 only possesses the semantic decoder, it reconstructs the stego image, which is designed to be a high-quality, natural-looking face that effectively conceals the secret information. The legitimate receiver undertakes a more complex task of mathematically recovering the original secret image through the noisy channel, rendering it inherently more sensitive to noise at $5~\text{dB}$.

\begin{table}[]
\centering
\caption{The steganalysis detection accuracy of XuNet \cite{xu2016structural}, where the detection accuracy is reported as the average over 10 independent runs with different random seeds.}
\label{tab:xunet}
\begin{tabular}{l c}
\toprule
Methods & Detection Accuracy (\%) \\
\midrule
Baluja\cite{baluja2019hiding} & 95.12\\
ISN\cite{lu2021large}& 56.23 \\
HiNet\cite{jing2021hinet} & 55.71 \\
AgentSemSteCom & 50.91 \\
\bottomrule
\end{tabular}
\end{table}

Figure \ref{evasdropper} presents the visual comparison of the recovered images for the legitimate receiver and the three eavesdropping scenarios at $\text{SNR}=10~\text{dB}$. The legitimate receiver achieves a faithful and high-fidelity reconstruction of the secret face. Eavesdropper 1 just recovers the stego image. Eavesdropper 2 with an incorrect digital token 856427, reconstructs with severe structural distortion and chaotic linear artifacts. The wrong perturbation mask and reference image leads to excessive sharpening, broken facial contours, and semantically incoherent details, where the original facial features are entirely overwhelmed by noise. Eavesdropper 3 recovers without any digital token, showing an over-smoothed artifacts, characterized by blurry and indistinct color patches. The absence of the token-determined latent initialization makes the diffusion model fail to recover accurate information in high-frequency domain.

To further evaluate the covertness of the proposed scheme against eavesdroppers, a classical deep learning-based steganalysis network XuNet \cite{xu2016structural} is employed to detect whether the transmitted images contains secret information. The network is tasked with distinguishing between the natural original images and the generated stego images, which are trained for 500 epochs. As shown in Table \ref{tab:xunet}, the traditional cover-based steganography approaches such as Baluja \cite{baluja2019hiding}, suffer from a high detection vulnerability with an accuracy of 95.12\%, and invertible neural network-based approaches like ISN \cite{lu2021large} and HiNet\cite{jing2021hinet} reduce the detection rate to 56.23 \% and 55.71\%, respectively. Notably, the proposed AgentSemSteCom scheme achieves a detection accuracy of 50.91\%, which reflects the generated stego images are statistically indistinguishable from natural image distributions, proving that AgentSemSteCom exhibits superior anti-steganalysis capabilities and can effectively confuse eavesdroppers in intellicise wireless networks.

\paragraph{The trade-off between security and covertness}
\begin{figure}
\centering
\includegraphics[width=0.9\linewidth]{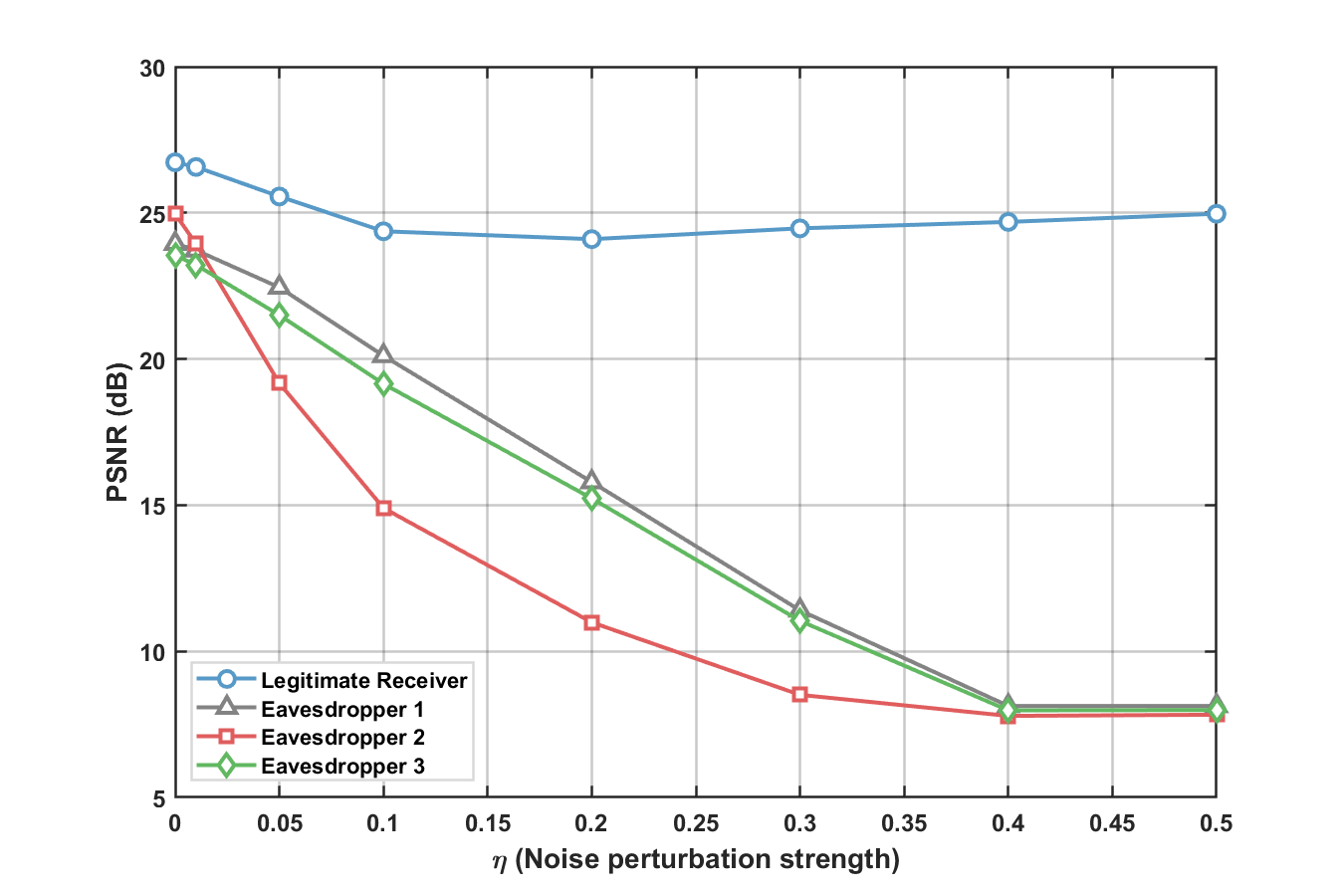}
\caption{Comparison of AgentSemSteCom with different noise perturbation strengths.}
\label{noiseFlip_matlab}
\end{figure}
\begin{figure}
\centering
\includegraphics[width=0.9\linewidth]{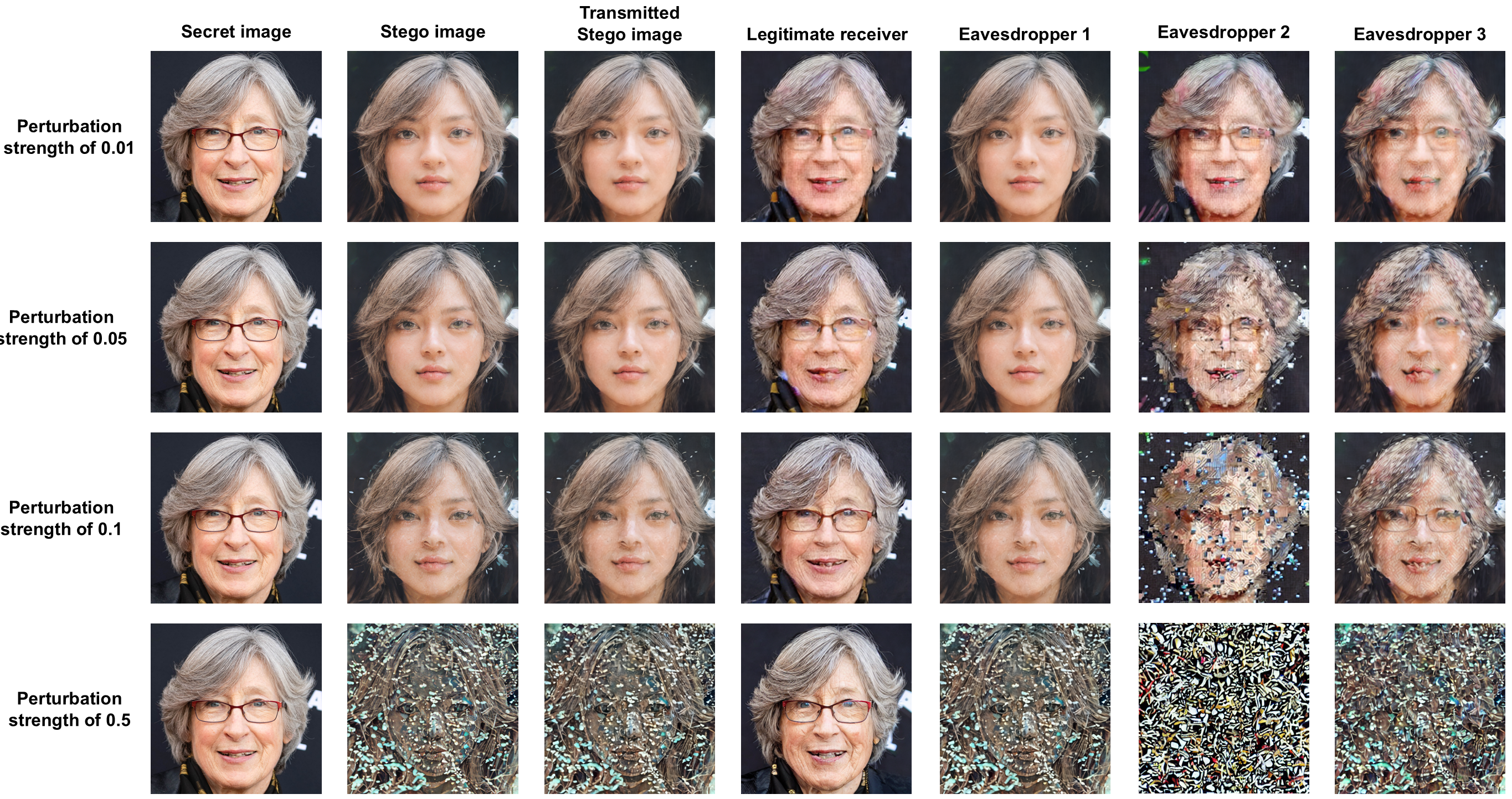}
\caption{Visualization of AgentSemSteCom with different noise perturbation strengths.}
\label{flipfigure}
\end{figure}
The noise perturbation strength $\eta$ determines the scale of latent noise components that are deterministically inversed. As illustrated in Figure \ref{noiseFlip_matlab}, an increase in this parameter significantly enhances the security of the steganography by widening the performance gap between the legitimate receiver and potential eavesdroppers. Specifically, as $\eta$ scales from $0.01$ to $0.5$, the PSNR for the eavesdropper 2 plummets from $23.94~\text{dB}$ to $7.83~\text{dB}$, indicating a total collapse of unauthorized recovery.
Figure \ref{flipfigure} shows the visualization result of the stego image and recovered results correspond to $\eta$ of $0.01$, $0.05$, $0.1$, and $0.5$ at $10~\text{dB}$ SNR. As $\eta$ increases, the reconstructed results of agentic AI-based eavesdroppers is highly disordered as the last column. However, this enhanced security is achieved at the cost of degraded visual naturalness. For the legitimate receiver shown in the third column, a large $\eta$ of $0.5$ induces obvious artifacts. Specifically, the high-frequency facial details, such as fine lines and wrinkles, are over-emphasized and transformed into unnaturally sharp contours. Meanwhile, the corresponding stego images in the first column show an engraved appearance, with diminished natural texture in skin regions, which could increase the suspicion risk of eavesdroppers. These observations indicate that although a large perturbation strength maximizes resistance against unauthorized recovery, it simultaneously degrades the covertness of stego image. Consequently, agentic receiver must adaptively select a moderate perturbation strength to strike balance between security and covertness, ensuring that the stego image is irrecoverable while preserving a visually natural appearance.


\section{Conclusions}
In this paper, the AgentSemSteCom scheme was proposed to address the security challenges in intellicise wireless networks. The framework integrates the autonomous semantic extraction, digital token-controlled reference generation, conditional diffusion model-based coverless steganography, JSCC-based semantic codec, and optional task-oriented enhancement modules. By leveraging agentic AI’s capabilities in perception and reasoning, the system eliminates the traditional dependence on private semantic keys and original cover images, thereby realizing a more secure and flexible ``invisible" encryption paradigm. Specifically, AgentSemSteCom handles the noising and sampling process within the latent space, where the coordination of public semantic keys and implicit structural features ensures controllable and high-fidelity generation. To guarantee the invertible of diffusion, we incorporated EDICT into the generative logic, effectively mitigating semantic drift caused by linearization errors in standard diffusion processes. Furthermore, a digital token-based security mechanism was implemented to provide dual-stage protection through deterministic noise initialization and latent perturbation, ensuring that only legitimate receivers can reconstruct the secret content. Experimental results on the open-source datasets demonstrated the superiority of the proposed AgentSemSteCom scheme in terms of pixel-level accuracy and structural consistency over the benchmark scheme. The autonomous invocation of task-oriented enhancement tools further highlighted the framework's adaptability to diverse task requirements and channel conditions. 

Future work will investigate step-efficient-denoising-based coverless steganography to meet stringent latency demands in high-speed mobility scenarios. Additionally, we aim to combine model compression techniques, such as knowledge distillation, quantization, and fine-tuning, to address computational and energy constraints inherent in space-air-ground integrated networks.

\bibliographystyle{IEEEtran}
\bibliography{ref}

\end{document}